\documentclass[aps,prx,showpacs,preprintnumbers,twocolumn,superscriptaddress,nofootinbib,10pt]{revtex4-2}
\usepackage{amsmath,amssymb}
\usepackage{bm}
\usepackage{tipa}
\usepackage{upgreek}
\usepackage{comment}
\usepackage{mathrsfs}
\usepackage{graphicx}
\usepackage{braket}
\usepackage{enumitem}
\usepackage{mathtools}
\usepackage{mathbbol}
\usepackage{gensymb}
\usepackage[normalem]{ulem}
\usepackage{color}
\usepackage[colorlinks,bookmarks=true,citecolor=blue,linkcolor=red,urlcolor=blue]{hyperref}
\usepackage{hyperref}

\usepackage{pifont}

\newcommand{\dpi}[1]{\frac{\text{d}#1}{2\pi}\text{ }}
\renewcommand{\d}[1]{\mathrm{d}#1\,}
\newcommand{\bbZ}{\mathbb{Z}}
\DeclareMathSymbol{\shortminus}{\mathbin}{AMSa}{"39}
\newcommand{\shortplus}{%
        \mathbin{%
            \raisebox{0.125ex}{\scalebox{0.5}{%
                \ensuremath{\boldsymbol{+}}%
            }}%
        }%
}
\usepackage{soul}

\newcommand{\Z}{\langle \sigma^z_0 \rangle }
\newcommand{\Zr}[1]{\langle #1 \vert \sigma^z_0 \vert 0_\shortminus \rangle }
\newcommand{\Zl}[1]{\langle 0_\shortminus \vert \sigma^z_0 \vert #1 \rangle }
\newcommand{\Zm}[2]{\langle #1 \vert \sigma^z_0 \vert #2 \rangle }

\begin{document}

\title{Real-time bubble nucleation and growth for false vacuum decay on the lattice}
	
	\author{Daan Maertens}
	\affiliation{Department of Physics and Astronomy, University of Ghent, Krijgslaan 281, 9000 Gent, Belgium}
	\author{Jutho Haegeman}
	\affiliation{Department of Physics and Astronomy, University of Ghent, Krijgslaan 281, 9000 Gent, Belgium}
    \author{Karel Van Acoleyen}
    \affiliation{Department of Physics and Astronomy, University of Ghent, Krijgslaan 281, 9000 Gent, Belgium}
    \affiliation{Department~of~Electronics~and~Information~Systems, University~of~Ghent,~Technologiepark-Zwijnaarde~126,~9052 Gent,~Belgium}

	\begin{abstract}
    {We revisit quantum false vacuum decay for the one-dimensional Ising model, focusing on the real-time nucleation and growth of true vacuum bubbles. Via matrix product state simulations, we demonstrate that for a wide range of parameters, the full time-dependent quantum state is well described by a Gaussian ansatz in terms of domain wall operators, with the
    associated vacuum bubble wave function evolving according to the linearized time-dependent variational principle. The emerging picture shows three different stages of evolution: an initial nucleation of small bubbles, followed by semi-classical bubble growth, which in turn is halted by the lattice phenomenon of Bloch oscillations. Furthermore, we find that the resonant bubble only plays a significant role in a certain region of parameter-space. However, when significant, it does lead to an approximately constant decay rate during the intermediate stage. Moreover, this rate is in quantitative agreement with the analytical result of Rutkevich (Phys. Rev. B {\bf 60}, 14525) for which we provide an independent derivation based on the Gaussian ansatz. 
}

    \end{abstract}
	
 	\maketitle

\tableofcontents

\section{Introduction}
It is possible that the current universe exists in a meta-stable state, a high energy excited state of the true ground state with an (extremely) long lifetime, known as the false vacuum \cite{Electroweak}. In such a scenario, the universe could transition to the true vacuum through a process known as false vacuum decay. This violent process could have far-reaching cosmological consequences, potentially altering the fundamental constants of nature and thus ending life as we know it. Not wanting to trigger such a catastrophe ourselves, we investigate false vacuum decay in the safer setting of a quantum spin chain.

The phenomenon of false vacuum decay was first studied in the context of phase transitions in classical thermodynamics. Foundational work by Becker and Döring \cite{BeckerDöring1935}, Frenkel \cite{Frenkel1939}, and Zeldovich \cite{Zeldovich1943} during the 1930s and 1940s, led to the development of what is now known as classical nucleation theory \cite{Kalikmanov2012}. The underlying assumptions and heuristics of this framework were later derived from first principles by Langer \cite{Langer} using methods of statistical physics. The ideas of Langer were extended to quantum field theory (QFT) in the works of Kobsarev, Okun and Voloshin \cite{Voloshin}, and of Coleman and Callan\cite{Coleman1,Coleman2}. The main result of Coleman and Callan's seminal work was the application of the path integral formalism to calculate the decay rate for a given QFT model, accurate to the leading order in $\hbar$, along with the first quantum correction. In particular, the method relies on solving the classical equations of motion in imaginary (Euclidian) time, i.e. computing the action of a certain instanton-like solution known as the ``bounce''. As a lattice counterpart to these QFT computations, Rutkevich \cite{Rutkevich}  presented
an analytic prediction of the false vacuum decay rate for the $d=1+1$ quantum Ising model. The central picture that emerges, both on the lattice and in the continuum, is that false vacuum decay is driven by the production (nucleation) of resonant bubbles of true vacuum with zero net energy cost, that will expand essentially in a classical manner, thereby balancing out the decrease in bulk potential energy with an increase in kinetic energy of the bubble walls. But note that these works focused exclusively on the asymptotic (large time) stationary regime.

There has been a recent surge of interest in the false vacuum decay phenomenon from the perspective of non-equilibrium quantum dynamics, investigating the full real-time character of this process. This is particularly interesting in light of the recent advances in the field of quantum simulation, increasingly bringing this phenomenon within experimental reach. Additionally, modern classical simulation methods such as tensor network or truncated Hamiltonian methods offer the possibility of studying false vacuum decay in real-time. Hereto, the false vacuum decay (or lack thereof) is studied as a quantum quench, starting from an initial state that closely resembles the false vacuum of the post-quench Hamiltonian. These approaches have been applied to a variety of quantum field theories or quantum lattice models in 1+1 dimensions, such as the paradigmatic $\lambda \varphi^4$ model \cite{DecayPhi4,Abel2025} and Ising spin chain \cite{Lagnese,Wilczek,BlochOscillations,Lencses2022}, as well as models with a richer spectrum of excited states, such as the tricritical Ising model \cite{Lencses2022}, the Potts model \cite{Potts} or neutral atom systems with Rydberg interactions \cite{NeutralAtoms}. One of the earliest tensor-network studies, Ref.~\onlinecite{Milsted}, investigated the real-time scattering of localized false vacuum bubbles on top of the true vacuum, in order to model the late-time aspects of the decay. Various other setups to study the real-time aspects of false vacuum decay have been proposed, e.g. the real-time evolution of initial configurations sampled from the full quantum state using classical equations of motion \cite{RealtimeDecay1} or the use of monitored or measurement-assisted evolution \cite{MonitoredIsing}. For quantum many-body systems, a rigorous theory of metastable states and their decay process was recently presented in \cite{Yin2025}. Experimental observation of false vacuum decay has also been realized recently, both in ferromagnetic superfluids \cite{FerroSuperFluid} and on a quantum annealer \cite{QuantumAnnealer}.

Previous studies have mainly focused on unraveling the signature of the (onset of) false vacuum decay through observables such as magnetization and two-point correlation functions. In this work, we aim to directly explore the bubble nucleation and expansion processes via a dedicated ansatz for the time-evolved state. This enables us to explore how the semi-classical picture of false vacuum decay fits into an entirely quantum setting, which we investigate for the transverse-field Ising spin chain with a longitudinal (integrability-breaking) field.

The structure of our work is as follows. In section \ref{sec:ModelQuench} we explain the essentials of the Ising spin chain and the quench protocol that we use to study the false vacuum decay. Section~\ref{sec:Ansatz} introduces the 
squeezed-state ansatz for the time-evolved state, parametrized by a quantity that we interpret as the bubble ``wave function'', and derive its dynamics using the time-dependent variational principle. In section~\ref{sec:Bubbledynamics}, we demonstrate the validity of our approach by comparing the squeezed-state ansatz with matrix product state simulations, and then analyze the bubble dynamics in real space and momentum space. We expand on the role of the resonant bubble by analyzing the bubble dynamics in energy space in section~\ref{sec:ResBubble}.  Finally, our conclusions and outlook are presented in section \ref{sec:Conclusions}.

\section{Model and Quench protocol} \label{sec:ModelQuench}
This section reviews the necessary details of the quantum Ising spin chain with transverse and longitudinal field.

\subsection{Quantum Ising chain and fermionic description}
In the thermodynamic limit, the Hamiltonian for the quantum Ising chain, with transverse field $h_\perp$ and longitudinal field $h_\parallel$, reads
\begin{equation} \label{eq:Hxz}
    \hat{H} = -\sum_{n\in \bbZ} \sigma^z_{n} \sigma^z_{n+1} - h_\perp \sum_{n\in\bbZ}  \sigma^x_n - h_\parallel \sum_{n\in\bbZ}  \sigma^z_n\,,
\end{equation}
where $\sigma_n^\alpha$ ($\alpha=x,y,z$) are the Pauli operators. When $h_\parallel=0$, this model is referred to as the transverse field Ising model (TFIM) and the Hamiltonian $\hat{H}_0$ has a $\bbZ_2$ symmetry, which is spontaneously broken for $\vert h_\perp \vert < 1 $. As a result, this phase exhibits long range order with two ordered ground states $\vert 0_{\shortplus}  \rangle$ and $\vert 0_\shortminus \rangle$, differentiated by their magnetization $ \langle 0_\pm \vert \sigma^z_0 \vert 0_\pm \rangle = \pm \Z = \pm(1-h_\perp^2)^{1/8}$. Via a Jordan-Wigner transformation, the TFIM is cast into a free fermion Hamiltonian
\begin{equation}\label{eq:H0}
\begin{split}
    \hat{H}_0 &= \sum_{n\in\bbZ} (\hat{c}_n^\dagger - \hat{c}_n)(\hat{c}_{n+1}^\dagger+\hat{c}_{n+1}) \\
    &\qquad+ h_\perp \sum_{n\in\bbZ}  (\hat{c}_n \hat{c}^\dagger_n - \hat{c}^\dagger_n \hat{c}_n)\\
    &= E_0 + \int_{-\pi}^{\pi} \dpi{k}  \omega(k) \hat{\gamma}^\dagger(k)\hat{\gamma}(k)\,,
\end{split}
\end{equation}
with ground state energy $E_0$ and dispersion relation
\begin{equation}
    \omega(k) = 2\sqrt{(h_\perp-\cos k)^2+\sin^2 k}\,.
\end{equation} 
The real-space fermionic creation and annihilation operators $(\hat{c}_n^\dagger, \hat{c}_n)$ are related to the momentum space operator $(\hat{\gamma}^\dagger(k),\hat{\gamma}(k))$ via a Bogoliubov and Fourier transform as outlined in Appendix~\ref{app:sec:Diagonalization}. We can further introduce real-space, but non-local, versions of the ladder operators via
\begin{align} \label{eq:dw_operator}
    \hat{b}_n &= \int_{-\pi}^{\pi} \dpi{k} \hat{\gamma}(k) e^{-ik/2}e^{ikn}\,,\\
    \hat{b}^\dagger_n &= \int_{-\pi}^{\pi} \dpi{k} \hat{\gamma}^\dagger(k) e^{ik/2}e^{-ikn}\,.
\end{align}
In the ordered phase, the operators $\hat{b}^\dagger_n$ and $\hat{b}_n$ respectively create and annihilate a domain wall (kink) centered at site $n$. The inclusion of the additional phase factor $e^{\pm ik/2}$ factor ensures that in the limit $h_\perp \to 0$, the $b$ operators become strictly local in terms of original spin operators.

After switching on $h_\parallel$, the $\bbZ_2$ symmetry of the model gets broken explicitly, removing the ground state degeneracy. In the limit $h_\parallel\rightarrow 0$, the state $\vert 0_\pm \rangle$ which aligns favorably with $h_\parallel$ becomes the ground state  (the \textit{true vacuum}) while the other state can be seen as a meta-stable state (the titular \textit{false vacuum}).

\subsection{Quench protocol and simulation methods}
In this manuscript, we study the following quench protocol for different particular values of $h_\perp$ (see also \cite{Wilczek}). We begin by preparing the initial state as the ground state of $\hat{H}_0$ with \textit{negative} magnetization, denoted as $\vert 0_\shortminus\rangle$. This state is subsequently evolved with $\hat{H}$ for $h_\parallel > 0$. For $h_\parallel$ sufficiently small, the initial state closely approximates the false vacuum for the post-quench Hamiltonian.

This quench is then studied using two different approaches. Our main strategy involves an approximate ansatz, namely a fermionic squeezed state ansatz first proposed in \cite{Bastianello_squeezed}, to which we apply the (linearized) time-dependent variational principle (TDVP). This ansatz, together with the resulting TDVP equation, are discussed at length in the next section. It has the advantage of offering a clear physical interpretation to the resulting dynamics, albeit being valid only in a particular region of the parameter space of $h_\perp$ and $h_\parallel$.

To assess the validity of this ansatz, we contrast its predictions with quasi-exact simulations of the dynamics using matrix product states (MPS) \cite{White1993,Verstraete2008,TangentSpaceMethods}. These simulations are performed using the formalism of infinite MPS (iMPS), using the open-source package MPSKit.jl \cite{MPSKit}. Ground states are obtained using the VUMPS algorithm \cite{VUMPS}, where the convergence tolerance is set to $\varepsilon=10^{-10}$. The time-evolution is carried out using the TDVP for MPS algorithm \cite{TangentSpaceMethods}, with a fixed time step of $\mathrm{d}t=0.01$. Throughout both the ground state and time evolution simulations, the bond dimension $D$ of the state is dynamically grown according to the algorithm outlined in Ref.~\onlinecite{VUMPS}, up to a maximal bond dimension $D_\text{max}=250$ unless stated otherwise, where during the process only Schmidt values larger than a cutoff $\eta$ are kept. By using two different values of $\eta$, we can verify the accuracy of the MPS simulations. For reasons of clarity, figures that plot a time-dependent quantity will only include subset of the MPS data, i.e. at coarser time steps than the simulation step size of $\mathrm{d}t=0.01$.

\section{Squeezed-state ansatz} \label{sec:Ansatz}
This section discusses the variational ansatz and its properties, as well as the physical interpretation of the variational parameters and their dynamics. 
\subsection{Gaussian ansatz}
Our main tool to study the quench protocol outlined above, is the (Gaussian) fermionic squeezed-state ansatz given as
\begin{equation} \label{eq:squeezed_ansatz}
    \vert \psi(t) \rangle = \frac{1}{\sqrt{N}}\exp \left( \int_0^\pi \dpi{k} K(k,t) \gamma^\dagger(k) \gamma^\dagger(-k) \right) \vert 0_\shortminus \rangle \,,
\end{equation}
where $\vert 0_\shortminus \rangle$ is the false vacuum, so we have the initial condition $K(k,0)=0$. Since the operators $\gamma^\dagger(k)$ create a domain wall with momentum $k$, we can interpret this state as an (incoherent) extensive superposition of bubbles with zero total momentum.

The exact same ansatz, but then starting from the true vacuum, was proposed in Ref.~\onlinecite{Bastianello_squeezed}, to study the dynamics of entanglement under quenches. In that case, the ansatz describes the superposition of pairs of confined domain wall excitations, a.k.a.\ mesons, which in this context we could also think of as small false vacuum bubbles.

Note that the form of the exponent allows us to extend $K(k,t)$ to negative momenta with $K(-k,t)=-K(k,t)$. The normalization $N$ of the state is given by
\begin{equation}
      N = \exp \left(L \int_0^\pi \dpi{k} \log[1+\vert K(k,t)\vert^2] \right)\,,
\end{equation}
where $L$ represents the formally infinite length (volume) of the system, and reflects the extensive nature of the ansatz. Furthermore, the squeezed state is readily expressed in real-space as 
\begin{equation}
     \vert \psi(t) \rangle =\frac{1}{\sqrt{N}}  \exp \left( \sum_{n_1>n_2} W(n_1-n_2,t) b_{n_1}^\dagger b_{n_2}^\dagger \right) \vert 0_\shortminus \rangle\,,
\end{equation}
with
\begin{equation}
    W(n,t) = \int_{-\pi}^\pi \dpi{k} K(k,t) e^{ikn}\,,
\end{equation}
where we have that $W(-n,t)= -W(n,t)$ due to the skew-symmetric nature of $K(k,t)$.

In the limit of small $K(k,t)$, the state can be understood perturbatively as
\begin{equation} \label{eq:squeezed_ansatz_perturbation}
     \vert \psi(t) \rangle \approx \frac{1}{\sqrt{N}} \left(\vert 0_\shortminus \rangle + \int_0^\pi \dpi{k} K(k,t) \vert k,-k\rangle + \dots \right)\,,
\end{equation}
with now
\begin{equation}
\begin{split}
    N &\approx \exp \left(L \int_0^\pi \dpi{k}\vert K(k,t)\vert^2 \right) \\ 
    &= \exp \left(L \sum_{n>0} \vert W(n,t)\vert^2 \right) \,,
\end{split}
\end{equation}
and where we introduced the notation
\begin{equation}
    \vert k_1, k_2 \rangle = \gamma^\dagger(k_1) \gamma^\dagger(k_2) \vert 0_\shortminus \rangle\,.
\end{equation}

\subsection{Bubble probability and decay rate}
\label{ss:decayrate}
The physical meaning of $W(n,t)$ and $K(k,t)$ becomes clear when considering the operator
\begin{equation}
    \hat{P}_{(0,n)} =  \hat{P}_\mathrm{even} b^\dagger_0 b_0 b_1 b^\dagger_1 b_2 b^\dagger_2 \dots b_{n-1} b^\dagger_{n-1} b^\dagger_n b_n \,,
\label{eq:proj}\end{equation}
which projects onto domain walls at site $0$ and $n$, no domain walls in between and contains a factor
\begin{equation}
    \hat{P}_\mathrm{even} = \frac{1}{2} \left(1+e^{i\pi\sum_{m<0} b_m^\dagger b_m} \right)\,,
\end{equation}
to project on an even number of domain walls to the left of site $0$. This ensures that all bubbles to the left of site zero have terminated, so that the domain wall on site zero has false vacuum to its left and true vacuum to its right. Taken together, the full projector $\hat{P}_{(0,n)}$ selects the states with a true vacuum bubble in between site $0$ and site $n$ and its expectation value $p_{(0,n)}= \braket{\hat{P}_{(0,n)}}$ therefore represents the probability for finding such bubble.

In the limit of small $W(n,t)$, we have that
\begin{align}
    \langle b_m b_n \rangle = \overline{\langle b^\dagger_n b^\dagger_m \rangle} &\approx W(m-n,t),\\
     \langle b_n b^\dagger_m \rangle &\approx \delta_{m,n} + \mathcal{O}(\vert W(m-n,t) \vert^2).
\end{align}
Using Wick's theorem, it can then be shown that for $W(n)$ small and decaying sufficiently fast in $n$ we get the following approximation:
\begin{equation} \label{eq:bubble_prob}
    p_{(0,n)} = \langle \hat{P}_{(0,n)} \rangle \approx \langle b^\dagger_0 b^\dagger_n \rangle \langle b_n b_0 \rangle \approx \vert W(n,t)\vert^2\,,
\end{equation}
so that $|W(n,t)|^2$ gives us the probability of finding a bubble between site $0$ and $n$. $W(n,t)$ and its Fourier transform $K(k,t)$ can thus be considered a \emph{bubble wave function or amplitude}.

The Gaussian ansatz also allows us to identify the false vacuum decay rate by considering the overlap between $\vert \psi(t) \rangle$ and the initial state $\vert 0_\shortminus \rangle$. The probability of the state not having decayed is given by:
\begin{equation}
    P_\mathrm{surv}(t) = \vert  \langle 0_\shortminus \vert \psi(t) \rangle \vert^2 = \frac{1}{N} := \exp \left(-\beta(t) \right) \,,
\end{equation}
where for small $W(n,t)$,
\begin{equation}
    \beta(t) \approx  L \sum_{n>0} \vert W(n,t)\vert^2\,.
\end{equation}
From this we find for the decay rate $\gamma(t)$ per site:
\begin{equation}
\gamma(t) = \frac{1}{L}\frac{\mathrm{d}\beta(t)}{\mathrm{d} t} \approx \frac{\mathrm{d}}{\mathrm{d} t} \left( \sum_{n>0} \vert W(n,t)\vert^2 \right)\,.\label{eq:decayrate}
\end{equation}
When $\gamma(t)\approx \gamma_0$ is approximately constant we have:  
\begin{equation}
    P_\mathrm{surv}(t) \sim  e^{-L\gamma_0 t}\,.
\end{equation}
In the following section we will show that for certain choices of $(h_\perp,h_\parallel)$ and for certain periods of time, we indeed find such behavior, and that the resulting $\gamma_0$ agrees well with the analytically computed decay rate of \cite{Rutkevich}. In section \ref{sec:ResBubble} we provide an alternative derivation of the analytical result, via Eq.~(\ref{eq:decayrate}).

\subsection{Time-dependent variational principle}
In Ref.~\cite{Bastianello_squeezed}, the time-dependence of $K(k,t)$ was obtained perturbatively, by expanding both the ansatz and the Hamiltonian and projecting everything onto the two-particle sector. Here, we follow a more general approach by applying the Dirac-Frenkel time-dependent variational principle \cite{Dirac1930,Frenkel1934,Kramer1981,Hackl2020}, which thus projects the linear Schr\"odinger dynamics onto the manifold of squeezed states. This yields a nonlinear differential equation for the time dependence of the variational parameters $K(k, t)$ or $W(n, t)$. In Appendix~\ref{app:sec:TDVP}, we work out that the TDVP equation for $K(k,t)$ is given by
\begin{equation} \label{eq:TDVP}
   \frac{i\Dot{K}(q,t)}{(1+\vert K_q \vert^2)^2} = \frac{2K(q,t)\omega(q)}{(1+\vert K_q \vert^2)^2} +h_\parallel \langle \gamma(q) \gamma(-q) \sigma^z_0 \rangle_{t,c}\,,
\end{equation}
where
\begin{equation} \label{eq:connected}
\begin{split}
  \langle \gamma(q) \gamma(-q) \sigma^z_0 \rangle_{t,c} &= \langle \gamma(q) \gamma(-q) \sigma^z_0 \rangle_t \\
  &\qquad- \langle \gamma(q) \gamma(-q) \rangle_t \langle \sigma^z_0 \rangle_t \,,
  \end{split}
\end{equation}
and the shorthand $\langle \cdot \rangle_t = \langle \psi(t) \vert \cdot \vert \psi(t) \rangle_t$ was used. Here, $\langle  \gamma(q) \gamma(-q) \sigma^z_0 \rangle_c$  is a complicated object to calculate in general. While the expression of $\sigma^z_0$ in terms of the fermionic ladder operators is known \cite{Jimbo}, its expectation value in the squeezed state  \eqref{eq:squeezed_ansatz} is not straightforward to obtain. It is this term which introduces nonlinear $K$ dependence, and gives rise to a Fredholm-type integral equation where $\dot{K}(q,t)$ depends on the value of $K(k,t)$ for all momenta $k \neq q$.

\begin{figure}
    \centering
    \includegraphics[width=\linewidth]{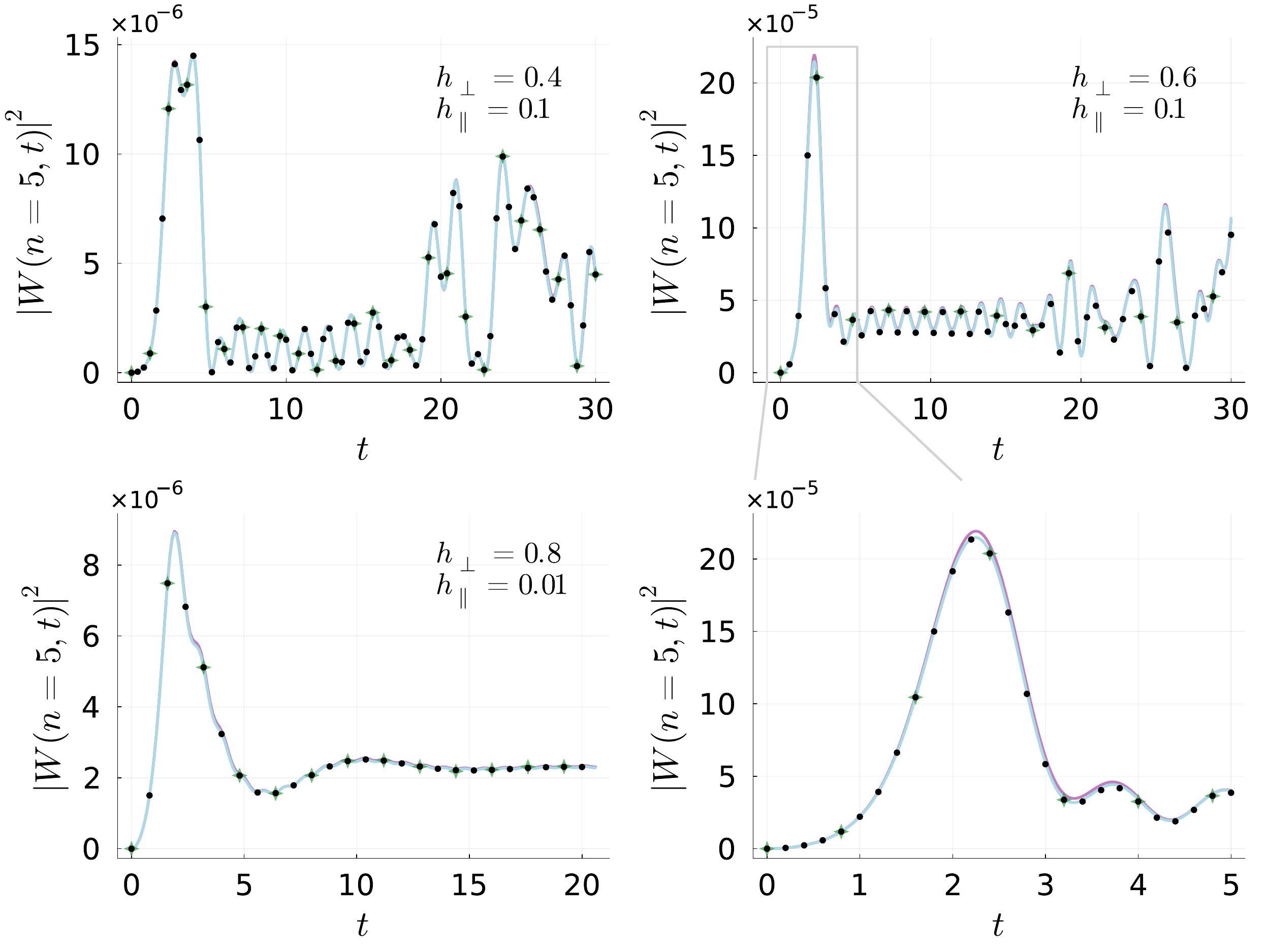}
    \caption{Comparison between $W(n,t)$ obtained by solving Eq.~\eqref{eq:TDVP_first_order} in full (blue line) and the same quantity extracted from MPS simulations, for different $h_\perp$ and $h_\parallel$ and a fixed $n=5$. Also shown is $W(n,t)$ obtained from solving Eq.~\eqref{eq:TDVP_first_order} without the $F$ and $G$ terms (purple line), corresponding to results from previous works \cite{Wilczek,Bastianello_squeezed}. MPS simulations were performed with truncation parameter $\eta = 10^{-5}$ (green stars) and $\eta = 10^{-6}$ (black dots). The inset (bottom right) displays the data
    for $h_\perp =0.6$ and $h_\parallel=0.1$ at higher time resolution and highlights the slight improvement gained by solving the linearized TDVP equation in full.}
    \label{fig:Wmps}
\end{figure}

\begin{widetext}
By treating $K(k,t)$ as a small parameter however, we can expand \eqref{eq:TDVP} up to first order to obtain the linearized TDVP equation
\begin{equation} \label{eq:TDVP_first_order}
     \begin{split}
       i\Dot{K}(q,t) =2K(q,t)\omega(q) + h_\parallel \biggl(\Zr{q,-q}  +\int_0^\pi \dpi{k} K(k,t) &\Zm{q,-q}{k,-k} - \Z L K(q,t) \\&+ \int_0^\pi \dpi{k}\overline{K(k,t)}\Zr{-k,k,q,-q}\biggr) \,,
    \end{split}
\end{equation}
where $ \langle \gamma(q) \gamma(-q) \sigma^z_0 \rangle_t$ was worked out using \eqref{eq:squeezed_ansatz_perturbation}. The divergent term $\Z LK(q,t)$, originating from the disconnected part in \eqref{eq:connected}, cancels out exactly with the hidden divergence in $\Zm{q,-q}{k,-k}$, leading to a well-defined finite expression.

The linear approximation is guaranteed to be valid at early times since $K(k,t)$ is zero initially, but can also be valid at later times as long as $\vert K(k,t)\vert$ remains small. Eq.~\eqref{eq:TDVP_first_order} can be further worked out with known expressions for the form factors $\langle q,-q \vert \sigma^z_0 \vert k,-k \rangle$ and other related matrix elements. The resulting expression is most instructive when written in real-space and reads
\begin{equation}  \label{eq:TDVP_first_order_realspace}
\begin{split}
        i\Dot{W}(n,t) =2 \sum_{n'\in\mathbb{Z}}T_{n-n'}W(n',t) -2h_\parallel\Z \vert n \vert W(n,t) - \frac{1}{2}  h_\parallel \Z\text{sign}(n)h_\perp^{\vert n \vert} \\ - h_\parallel\Z \sum_{n'\in \mathbb{Z}} \biggl( G_{n,n'}W(n',t) + F_{n,n'}\overline{W(n',t)} \biggr) \,,
\end{split}
\end{equation}
where the definitions of the newly introduced functions $T$, $G$ and $F$ were omitted for brevity and can be found in appendix \ref{app:sec:TDVP_solution}. Here, the first two terms in the right hand side correspond to the usual Schr\"{o}dinger equation with lattice kinetic operator $T_{n-n'}$ and a linear, anti-confining potential $-2h_\parallel\Z \vert n \vert$. The third term does not contain $W$ and stems from the matrix element $\Zr{q,-q}$ that enables transitions from the zero bubble state to states with multiple bubbles. As such, this source term is the driving force behind bubble formation.  The exponential suppression in $n$, means that the bubble formation is only significant at short distances. Finally, the last two terms represent quantum corrections that only affect small bubbles, due to the exponential decay of $G_{n,n'}$ and $F_{n,n'}$ in both $\vert n\vert$ and $\vert n'\vert $.
\end{widetext}
In the limit of small $h_\perp$ and $h_\parallel$, $G_{n,n'}$ and $F_{n,n'}$ become negligible and the solution of equation \eqref{eq:TDVP_first_order_realspace} reduces to the form previously reported in Refs.~\onlinecite{Rutkevich,Bastianello_squeezed,Wilczek}. Notice that the equations in these works were derived by specializing exclusively to the sectors containing zero and two domain walls; we refer again to appendix \ref{app:sec:TDVP_solution} for details. Beyond the small $h_\perp$ and $h_\parallel$ limit, we find that the numerical solution of \eqref{eq:TDVP_first_order_realspace} shows a slight improved agreement with numerical data compared to aforementioned previous works, as demonstrated in the inset plot of figure \ref{fig:Wmps}.

\section{Bubble dynamics} \label{sec:Bubbledynamics}

In this section, we study the bubble dynamics by interpreting the solution $W(n,t)$ of the (linearized) TDVP equation and contrast it with the corresponding MPS simulation of the same quench. 

\subsection{Ansatz verification and deviations}

We can identify two main sources of potential errors in our squeezed state approach. Firstly, there is the loss of Gaussianity, where the true time-evolving state deviates from being captured by the manifold of squeezed states. This is closely related to the projection error in the TDVP, originating from projecting the evolution back onto (the tangent space of) the variational manifold. Secondly, there are potential errors from linearizing the TDVP, as the first order equation can be expected to be valid only for $\vert K(k,t) \vert$ uniformly small in $k$.

To verify the predictions of the squeezed state ansatz and its evolution in the current setting, we will contrast it with MPS simulations. While we could simply compare predictions for e.g.\ time-dependent expectation values of local observables, we go down a different route and reconstruct $(Wn,t)$ directly form the MPS state.

For an associated MPS approximation to the time-evolving state, we can assess its deviation from Gaussianity by calculating the correlation matrix
\begin{equation}
    \Gamma_{n,m}(t) = \begin{pmatrix}
         \langle \hat{c}_n\hat{c}^\dagger_m\rangle & \langle \hat{c}_n\hat{c}_m \rangle \\ \langle \hat{c}^\dagger_n\hat{c}^\dagger_m \rangle & \langle \hat{c}^\dagger_n\hat{c}_m \rangle
    \end{pmatrix}\,.
\end{equation}
Because of translation invariance, the correlation matrix only depends on $m - n$, and we can define its Fourier transform
\begin{equation}
    \Gamma(k,t)=\sum_n e^{-ikn} \Gamma_{0,n}\,.
\end{equation}
For a true Gaussian state, the eigenvalues of $\Gamma(k,t)$ should be exactly 0 and 1, i.e.\ $\Gamma(k,t)$ has rank 1 and satisfies $\det \Gamma(k,t)=0$. Furthermore, in Appendix~\ref{app:sec:structfactor}, we show that $\Gamma(k,t)$ takes the form
\begin{align}
    \Gamma(k,t)   
   &= U_{\theta_{\shortminus k}} 
    \begin{pmatrix}
        \frac{1}{1+\vert K(k,t)\vert^2}  & \frac{K(k,t)}{1+\vert K(k,t)\vert^2}  \\ \frac{\overline{K(k,t)}}{1+\vert K(k,t)\vert^2}  & \frac{\vert K(k,t) \vert^2}{1+\vert K(k,t)\vert^2} 
    \end{pmatrix}  U_{\theta_{\shortminus k}}^\dagger \,,
\end{align}
in terms of the parameters $K(k,t)$ of the squeezed state ansatz and the Bogoliubov matrix $U_{\theta_{k}}$ that relates the original creation and annihilation operators to the $\gamma_k$ and $\gamma_k^\dagger$ operators. By computing $\Gamma(k,t)$ from the time-dependent MPS, we can thus invert this relation and determine the bubble wave function $K(k,t)$, and in turn $W(n,t)$, directly from MPS simulations, provided that the relation $\det \Gamma(k,t) \approx 0$ is satisfied.

\begin{figure}
    \centering
    \includegraphics[width=\linewidth]{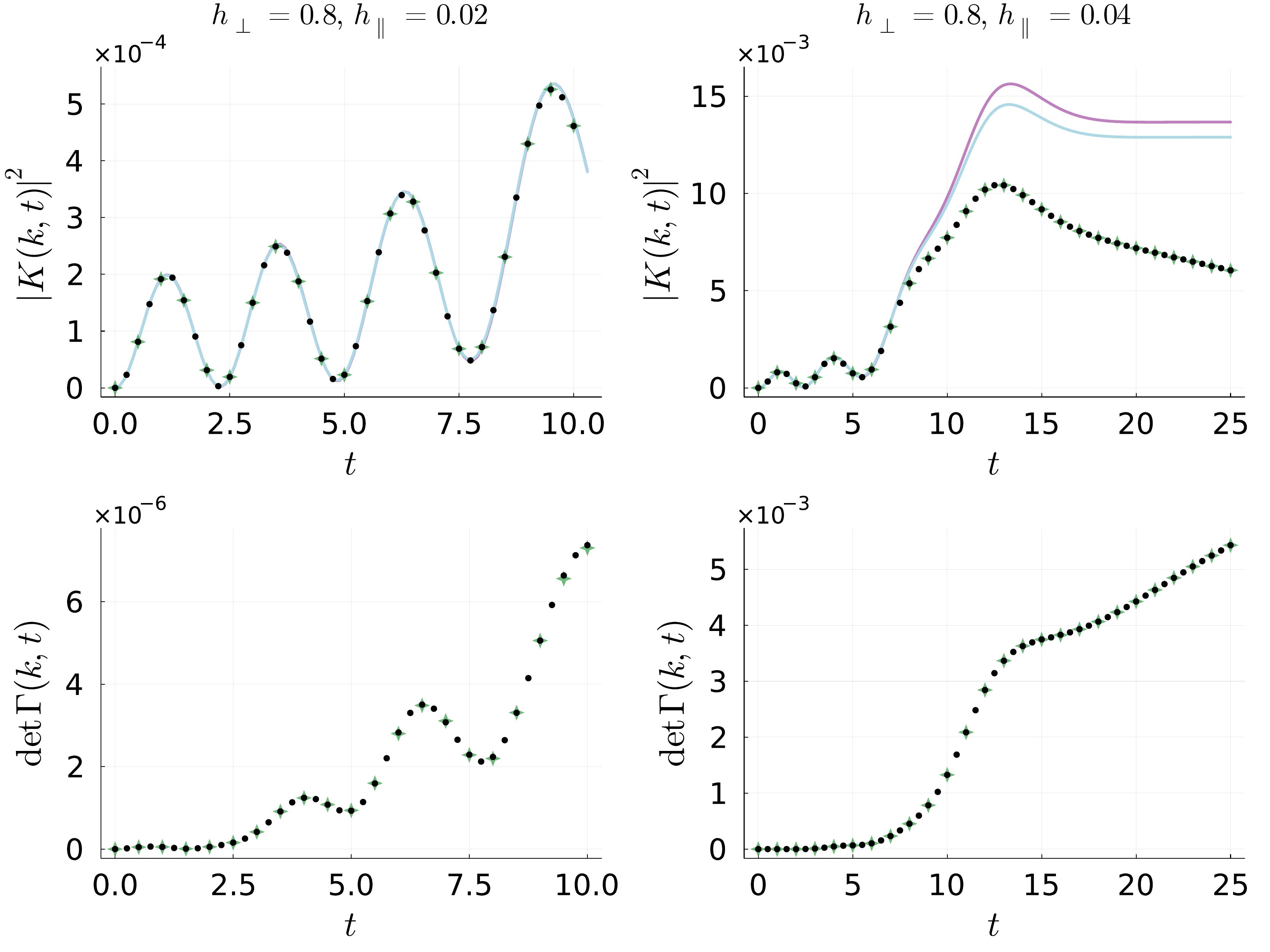}
    \caption{Comparison between the solution of the linearized TDVP equation Eq.~\eqref{eq:TDVP_first_order} and $K(k,t)$ obtained from MPS simulations (top row) and non-Gaussianity measure $\det \Gamma(k, t)$ of the MPS state (bottom row), for momentum $k=0.05\pi$ with the indicated couplings. MPS data are shown for $\eta =10^{-5}$ (green stars) and $\eta = \times 10^{-6}$ (black dots). The blue line depicts the full TDVP solution of Eq.~\eqref{eq:TDVP_first_order} while the purple line was obtained by solving the equation without the subdominant $F$ and $G$ terms.}
    \label{fig:GaussianDeviations}
\end{figure}

A comparison between $W(n,t)$ as obtained from solving Eq.~\eqref{eq:TDVP_first_order_realspace} and from the MPS simulations is shown in Figure~\ref{fig:Wmps} for various $h_\parallel$ and $h_\perp$ at a fixed position $n$. We find excellent agreement at all times for a wide range of $h_\perp$ and $h_\parallel$. However, for large values of $h_\parallel$ and $h_\perp\rightarrow 1$, the numerical solution of the linearized TDVP equation remains accurate only at early times. The discrepancies at later times may be due to higher order contributions to the TDVP equation, or from the breakdown of Gaussianity. Figure \ref{fig:GaussianDeviations} shows that deviations from Gaussianity become significant around the same time as the discrepancies set in, so that both effects are probably at play. This makes sense from a physical standpoint. The breakdown of the linearized regime implies large bubble densities, with non-vanishing inter-bubble interactions that are likely not captured by the Gaussian ansatz. 

Note also that, both Figures~\ref{fig:Wmps} and \ref{fig:GaussianDeviations} also display the solution of Eq.~\eqref{eq:TDVP_first_order_realspace} without $F$ and $G$ terms, i.e.\ according the dynamics as described in Refs.~\onlinecite{Rutkevich,Bastianello_squeezed,Wilczek} (purple lines). We find a minor difference with the solution of the full (linearized) TDVP equation, albeit with a slightly larger discrepancy with the MPS data (see in particular the inset of Fig. \ref{fig:Wmps}). 

Finally note that in the appendix \ref{app:sec:magnetization} we also show the verification of the squeezed state ansatz in terms of its predictions for the evolution of the local magnetizations $\braket{\sigma^x_0}_t$ and $\braket{\sigma^z_0}_t$.   

\subsection{Bubble nucleation and growth}

\begin{figure*}
    \centering
    \includegraphics[width=\linewidth]{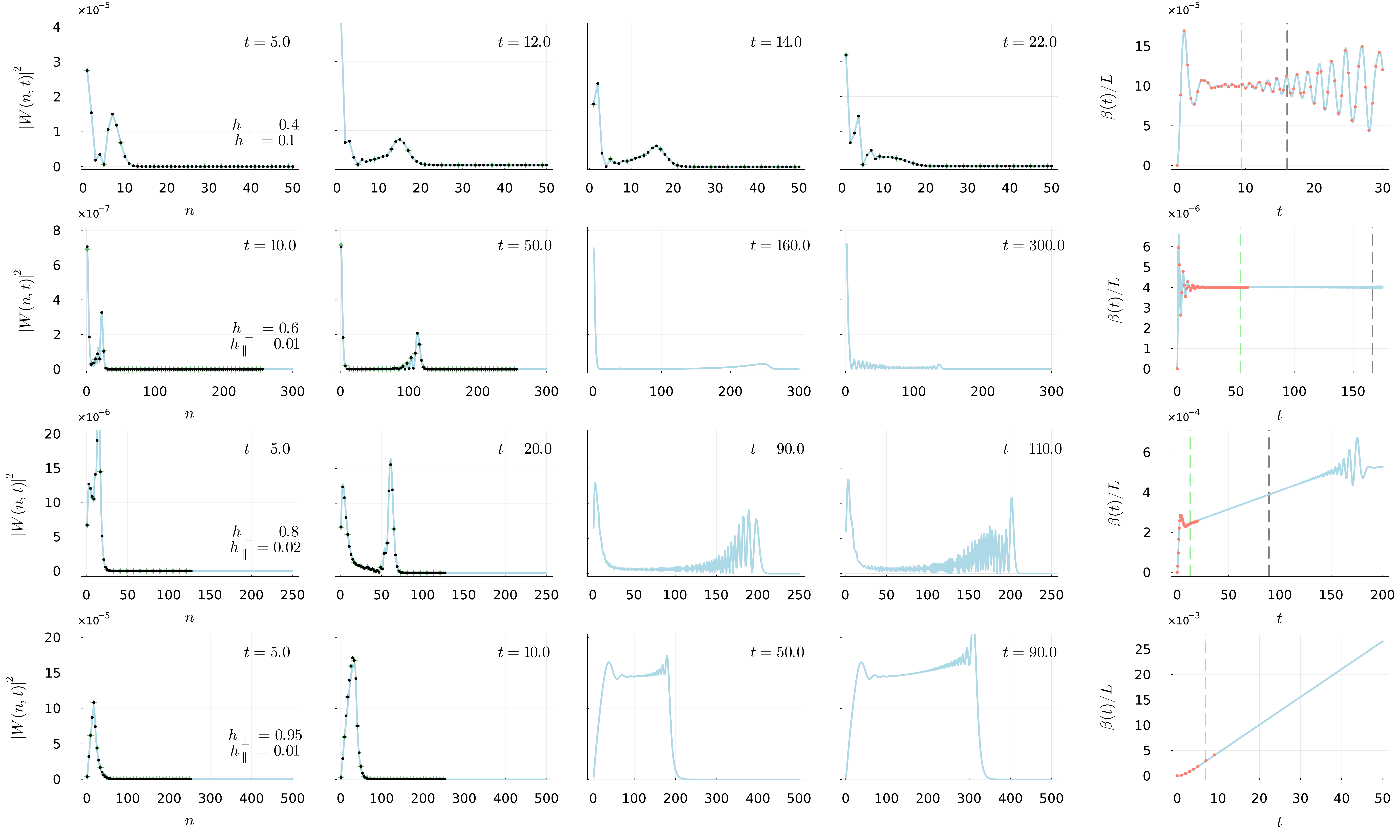}
    \caption{The observed real-space bubble dynamics. The first four columns show snapshot of the bubble wave function $W(n,t)$ at representative times for the parameters $h_\perp=0.4$, $h_\parallel=0.1$ (first row), $h_\perp=0.6$, $h_\parallel=0.01$ (second row), $h_\perp=0.8$, $h_\parallel=0.02$ (third row) and $h_\perp=0.95$, $h_\parallel=0.01$ (fourth row). MPS data is shown for $\eta=10^{-5}$ (green stars) and $\eta=10^{-6}$ (black dots), and the blue line indicates the full TDVP solution. For the $h_\perp=0.4,0.8,0.95$ simulations $D_{\text{max}}=250$, while for $h_\perp=0.6$, $D_{\text{max}}=500$. In the last column, the logarithmic overlap per volume $\beta(t)/L=-L^{-1}\ln \vert \langle 0 \vert \psi(t) \rangle \vert^2$ is shown for the same data. Here, the red dots represent the overlap computed directly from the MPS simulation with the highest accuracy, and convergence in $\eta$ was checked. The analytic time scales (see main text) for the onset of linear growth, $t_{\mathrm{lin}}$ (green dashed line), and Bloch oscillations, $t_{\mathrm{Bloch}}$ (grey dashed line) are indicated whenever they fall within the plotted range.}
    \label{fig:evolutiontypes}
\end{figure*}

Having established the validity of the squeezed state ansatz, we can now gain valuable insights into the decay process by analyzing the function $W(n,t)$. Indeed, by virtue of Eq.~\eqref{eq:bubble_prob}, we can interpret $\vert W(n,t) \vert ^2$ as the probability of finding a true vacuum bubble of length $n$ at time $t$ (in the regime where it is sufficiently small). Hence, this quantity enables us to track the bubble nucleation and growth process in real time.

Figure~\ref{fig:evolutiontypes} presents successive snapshots of $W(n,t)$ at representative times for four different choices of the parameters $(h_\perp, h_\parallel)$. Bubbles are generated by the source term (inhomogeneous term) in the (linearized) TDVP equation, i.e.\ the last term on the first line in Eq.~\eqref{eq:TDVP_first_order_realspace}. This term is exponentially suppressed away from the origin by a factor $h_\perp^n$. The initial peak that is generated starts to evolve to larger bubble lengths, while a static peak near the origin remains, such that $W(n,t)$ displays a bimodal distribution after some time (that is, until Bloch oscillations set it).

This behaviour can easily be understood from the structure of the (linearized) TDVP equation, which enables us to decompose $W(n,t)$ as
\begin{equation}
    W(n,t) = W^*(n) + W_{\text{dyn}}(n,t) \,.
\end{equation}
Here, the static part $W^*(n)$ corresponds to the solution of the (real linear) equation obtained by setting the right hand side of Eq.~\eqref{eq:TDVP_first_order_realspace} equal to zero. The dynamic part $ W^{\text{dyn}}(n,t)$ is governed by the (real) linear first order differential equation obtained by omitting the inhomogeneous source term from Eq.~\eqref{eq:TDVP_first_order_realspace}. The initial condition $W(n,0)=0$ translates into $ W^{\text{dyn}}(n,0) = - W^*(n)$. Since $G$ (and $F$) are also exponentially suppressed away from the origin, $ W^{\text{dyn}}(n,t)$ really adheres to the Schr\"odinger equation with kinetic energy operator $T$ and the linearly decreasing potential, once it is localized away from the origin. The evolution of this dynamical peak in the bubble distribution therefore reflects the prototypical behavior of growing bubbles that convert the excess energy from having a larger region of true vacuum into extra kinetic energy and momentum in the domain walls.

However, at some later time, the dynamic peak reaches a maximum distance $d_{\text{max}}$ and reverses direction, due to the well-known phenomenon of Bloch oscillations. These Bloch oscillations are caused by the boundedness of the lattice kinetic operator in \eqref{eq:TDVP_first_order} and are thus a pure lattice effect. This behavior is visible in the first three parameter configurations of Figure~\ref{fig:evolutiontypes}, where for the first configuration $(h_\perp=0.4, h_\parallel = 0.1)$ the Bloch oscillations kick in shortly after the appearance of the dynamical peak.

\begin{figure}
    \centering
    \includegraphics[width=\linewidth]{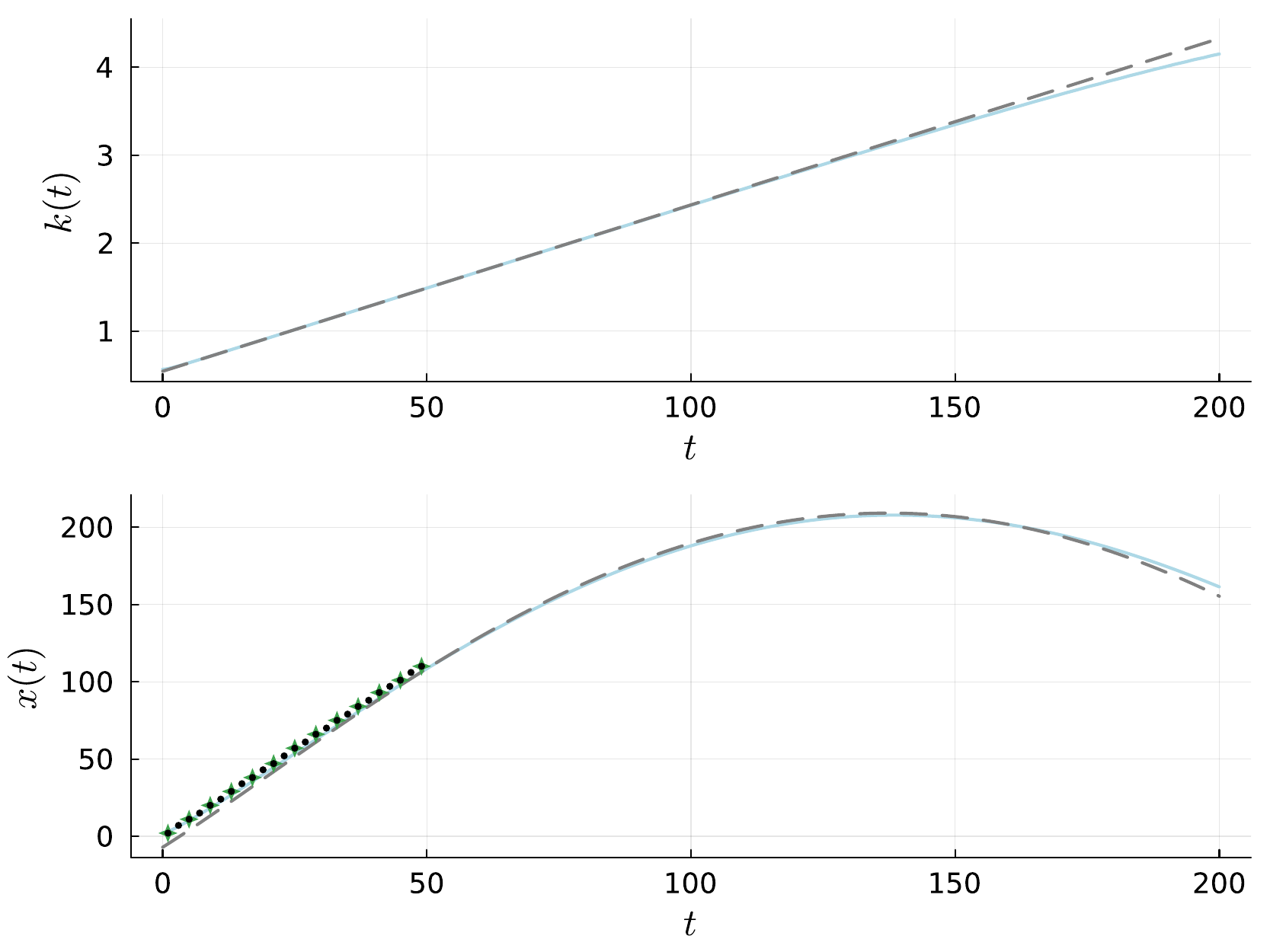}
    \caption{Semiclassical position $x(t)$ and momentum $k(t)$ of the dynamical part of $W(n,t)$ as a function of time for $h_\perp=0.6$ and $h_\parallel=0.01$. For the TDVP solution (blue line), $x(t)$ and $k(t)$ are computed as the average position $\langle x\rangle $ and momentum $\langle k \rangle$ w.r.t.\ the distribution $\vert W^{\text{dyn}}(n,t) \vert^2$ and $\vert K_\mathrm{dyn}(k,t) \vert^2$ respectively. MPS data are shown for $\eta=10^{-6}$ (black dots) and $\eta=10^{-5}$ (green stars), both with $D_\mathrm{max}=500$, and were obtained by tracking the position of the peak in $W^{\text{dyn}}(n,t)$. The grey dashed line corresponds to the solution of the classical Hamiltonian, matched with the TDVP solution at $t=50$.}
    \label{fig:semiclassical}
\end{figure}

The maximum bubble size $d_{\text{max}}$ and the associated time scale can be estimated by working out the equations of motion of the classical Hamiltonian (see also \cite{BlochOscillations,Rutkevich2008}),
\begin{equation}
    H_{\text{clas}} = 2\omega(k) - 2 h_\parallel \Z \vert x \vert \,.
\end{equation}
with $\omega(k)$ the $2\pi$-periodic dispersion relation. This yields
\begin{equation} \label{eq:EQM_Hclas}
\begin{split}
    \dot{x} &= 2\frac{\d\omega}{\d k}\,, \\
    \dot{k} &= 2 h_\parallel \Z \mathrm{sign}(x) \,,
\end{split}
\end{equation}
which for positive $x(0)$, has the solution
\begin{equation} \label{eq:Sol_Hclas}
    \begin{split}
        k(t) &= 2 h_\parallel \Z (t-t_0) + k_0 \,,\\
        x(t) &= \frac{1}{h_\parallel \Z} ( \omega(k(t))-\omega(k_0) + x_0\,.
    \end{split}
\end{equation}
Indeed, fitting these solutions (i.e.\ $x_0$ and $k_0$) to the dynamics of the first moment of $W(n,t)$ and $K(k,t)$\footnote{Because $K(k,t)$ is an odd function on $[-\pi,+\pi]$, its first moment is technically zero. Furthermore, $k$ is not a well-defined variable on the periodic Brillouin zone $[-\pi,+\pi]$, so that its average could never grow beyond $k=\pi$. We work around these issues by first defining an alternative Fourier transform $\tilde{K}(k,t)\propto \sum_{n > 0} W(n,t) \exp(-i k n)$ that only takes the physically relevant region $n>0$ into account, and then defining $\langle k \rangle(t) = \int_I k \lvert \tilde{K}(k,t)\rvert^2 \mathrm{d} k$ for a shifted integration interval $I$ of length $2\pi$ that is centered around the peak of the unimodal distribution $\lvert\tilde{K}(k,t)\rvert^2$.} shows excellent agreement, as illustrated in Figure~\ref{fig:semiclassical}.

From \eqref{eq:EQM_Hclas} and the form of $\omega(k)$, we gather that $x(t)$ attains a maximum and $\dot{x}$ reverses sign for $k(t)= \pi$, or thus at time
\begin{equation}
    t = t_0+\frac{\pi-k_0}{2 h_\parallel \Z}.
\end{equation}
Assuming that the initial momentum $k_0$ is sufficiently small to be ignored, we can define
\begin{equation}
    t_{\text{Bloch}} = \frac{\pi}{2 h_\parallel \Z}\,,
\end{equation}
as the characteristic time scale at which the Bloch oscillations occur. Correspondingly we find a maximal bubble length:  \begin{equation} \label{eq:dmax}
    d_{\text{max}}= x_0+ \frac{\omega(\pi)-\omega(k_0)}{h_\parallel \Z}\approx\frac{4h_\perp}{h_\parallel \Z}\,.
\end{equation}

\subsection{Total bubble probability and decay rate}
As discussed in Subsection~\ref{ss:decayrate}, the false vacuum survival probability can be related to the total bubble probability $\sum_{n>0} \lvert W(n,t)\rvert^2$. In the regime where the dynamical part $W^{\text{dyn}}(n,t)$ is well away from the origin and thus evolves according to a Schr\"odinger equation, the total probability associated with it is conserved, i.e.\
\begin{equation}
    \frac{\mathrm{d}}{\mathrm{d} t} \vert W^{\text{dyn}}(n,t) \vert^2 = 0\,.
\end{equation}
Hence, an increase in $\sum_{n>0} \lvert W(n,t)\rvert^2$ needs to originate from the remaining overlap between $W^{\text{dyn}}(n,t)$ and the static part $W^*(n)$, i.e.\
\begin{equation}
    \frac{\mathrm{d}}{\mathrm{d} t} \sum_{n>0} \vert W(n,t) \vert^2 = \frac{\mathrm{d}}{\mathrm{d} t}  2 \mathrm{Re}\left( \sum_{n>0} \overline{W^*(n)} W^{\text{dyn}}(n,t)\right)\,.
\end{equation}

For the last two parameter configurations of Figure~\ref{fig:evolutiontypes} we indeed observe the emergence of a large tail in between the static and dynamic peaks and correspondingly an approximately linearly rising $\beta(t)/L = \sum_{n>0} \lvert W(n,t)\rvert^2$, resulting in a constant false vacuum decay rate $\gamma_0 = \frac{1}{L} \frac{\mathrm{d} \beta(t)}{\mathrm{d} t}$, as shown in the right column of the figure. In the following section we will show that in the limit $h_\parallel \rightarrow 0$, one expects a constant decay rate for times $t_{\mathrm{lin}}\lesssim t \lesssim t_{\mathrm{Bloch}}$, with $t_{\mathrm{lin}}=|\ln(h_\perp)|/(h_z \Z)$. In our simulations we find that this first time scale $t_{\mathrm{lin}}$, which sets the onset of linear growth, matches reasonably well with the time at which the static and dynamic peak separate. For the first two parameter configurations of Figure~\ref{fig:evolutiontypes} we observe no rising of $\beta(t)/L$ after $t_{\mathrm{lin}}$. This is consistent with the analytical result for the decay rate (\ref{eq:anal_decay_rate}), which is exponentially small in those cases. In real-space this manifests itself in a vanishing tail between the two peaks of the bubble wave-function $W(n,t)$.

\section{Dynamics in terms of bubble eigenstates} \label{sec:ResBubble}

We now resolve the bubble nucleation and growth in the space of bubble eigenstates, in order to establish the precise role that is played by the resonant $E=0$ bubble. It will also allow us to provide an alternative derivation of the decay rate result of \cite{Rutkevich} and clarify its regime of validity. 

Following \cite{Rutkevich,Wilczek} we define the bubble eigenstates $\phi_l(n)$ as the stationary solutions of the Schr\"{o}dinger equation that arises from the first two terms---the kinetic-energy term and the anti-confining potential term---in the right-hand side of Eq. (\ref{eq:TDVP_first_order_realspace}): \begin{equation} \label{eq:eigprob_n}
  \sum_{n'\in\mathbb{Z}2}T_{n-n'}\phi_l(n') - 2 h_\parallel\Z \vert n \vert \phi_l(n)=  E_l\phi_l(n)\,,
\end{equation}
with the constraint $\phi_l(n)=-\phi_l(-n)$, $l$ the discrete label of the particular eigenstate and with the normalization convention: \begin{equation} \label{eq:norml}
\sum_{n>0} \overline{\phi_l(n)} \phi_{l'}(n) = \delta_{l,l'} \,.
\end{equation}  

\begin{figure}
    \centering
    \includegraphics[width=\linewidth]{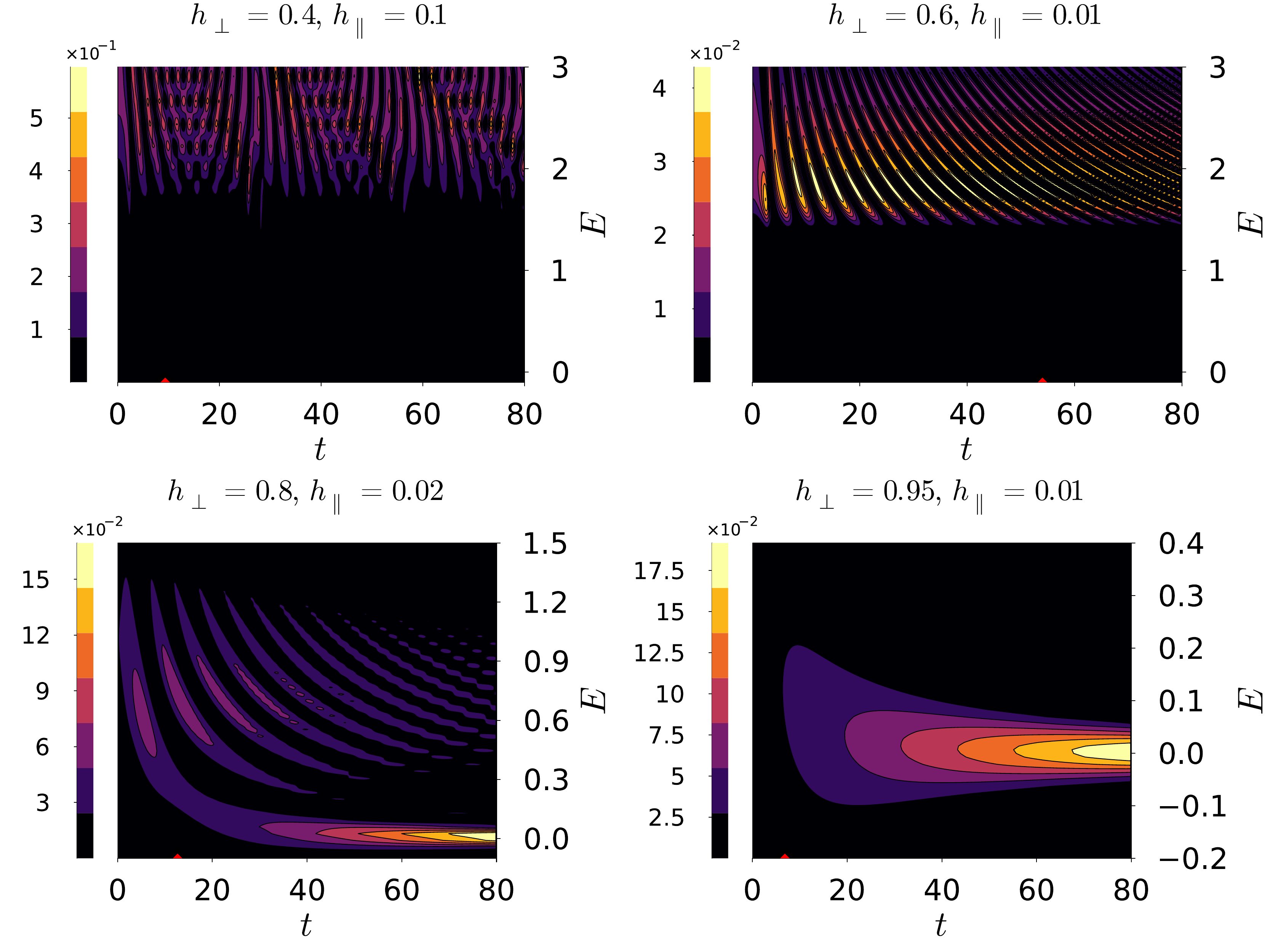}
    \caption{Contour plot of the (non) resonant behavior around $E\approx 0$ of $\vert c_E(t) \vert^2 / (\sum_E \vert c_E(t) \vert^2)$, as calculated from the (full) TDVP solution by projecting $W(n,t)$ onto $\phi_l(n)$, for various values of $h_\perp$ and $h_\parallel$. The time for the onset of linear growth $t_{\mathrm{lin}}$ is indicated with a red triangle on the time axis.}
    \label{fig:lspacedynamics}
\end{figure}

In Fig.\ \ref{fig:lspacedynamics} we show the dynamical evolution of the eigenstate wave function,
\begin{equation}
c_l(t)=\sum_{n>0}\overline{\phi_l(n)}W(n,t)\,,
\end{equation}
for the same parameter configurations as in Fig.\ \ref{fig:evolutiontypes}. As discussed above, the last two parameter configurations display a non-vanishing constant decay rate for $t\gtrsim t_{\mathrm{lin}}$, which in real-space manifests itself as a delocalized wave-function $W(n,t)$. As can be seen in Fig. \ref{fig:evolutiontypes}, in the bubble eigenstate space this corresponds to resonant behaviour, with the eigenstate wave function getting more and more localized around the resonant $E_l=0$ bubble.

To understand this behaviour more qualitatively we proceed by writing down the evolution induced by Eq.(\ref{eq:TDVP_first_order_realspace}) with the source term (third term) included, but omitting the subdominant fourth and fifth terms. This yields the solution
\begin{equation} \label{eq:cl}
 c_l(t) = \left(e^{-iE_l t}-1 \right) \frac{V_l}{E_l} = c_l^\text{dyn}(t) - c_l^*\,,
\end{equation}
where $c_l^\text{dyn}(t)$ and $c_l^*$ correspond to the eigenbasis expansion of the dynamic part $W^{\text{dyn}}(t)$ and static part $W^*$ of the bubble wave function, and where furthermore
\begin{equation} \label{eq:Vl}
    V_l =  -\frac{1}{2}h_\parallel\Z  \sum_{n>0} \phi_l(n) h_\perp^n \,.
\end{equation}

For the total bubble probability we then find:
\begin{eqnarray}
\frac{\beta(t)}{L}&=&\sum_{n>0} |W(n,t)|^2\nonumber\\
&=&\sum_{l} |c_l(t)|^2\nonumber\\
&=&4\sum_l \frac{V_l^2}{E_l^2} \sin^2\left(\frac{E_l t}{2}\right)\,,
\label{eq:step1}
\end{eqnarray}
where we anticipate a real-valued $V_l$. This equation suggests an approximate Fermi's golden rule, that is to say, for a discrete rather than a continuous spectrum. To make this explicit, we follow \cite{Rutkevich} and write down an approximation for $V_l$ at (near-)resonant energies $E_l\approx 0$, valid in the limit $h_\parallel\rightarrow 0$:
\begin{equation}
    V_l\approx \frac{1}{3} h_{\parallel}\braket{\sigma^z_0} 
    \exp\left(-\frac{q +E_l \ln h_{\perp}}{2 h_{\parallel} \Z} \right)\,,
    \label{eq:Vlapp}
\end{equation}
with
\begin{align}
E_l &\approx \frac{f(\pi)}{\pi}-2 h_{\parallel}\braket{\sigma^z_0} l\,,\label{eq:specapp}\\
q &= \vert f(i \ln(h_{\perp}) \vert \,, \\
f(\theta) &= 2\int^\theta_0\!d\alpha\, \omega(\alpha)\,.
\end{align} 
Our expression (\ref{eq:Vlapp}) differs in the pre-factor compared to \cite{Rutkevich} due to our different normalization (\ref{eq:norml}) and our non-extensive definition of $V_l$. Note also that, in addition to the $V_{E=0}$ result of \cite{Rutkevich}, we have included the dominant $E_l$-dependence around $E_l=0$.

For small enough values of the energy spacing $\Delta E=2 h_{\parallel} \Z$, the energy spectrum around $E_l=0$ becomes dense and we can turn the sum over $l$ into an integral over $E$,
\begin{equation}
    \sum_{l} |c_l(t)|^2 \approx \frac{2}{h_{\parallel} \Z} \int \frac{V(E)^2}{E^2} \sin^2\left(\frac{Et}{2}\right) \d E \,.
\end{equation}
Validity of this approximation requires the spectrum to be dense in region where the $\mathrm{sinc}^2$ factor has its dominant support, which amounts to the condition $t\lesssim \frac{\pi}{2 h_{\parallel}\braket{\sigma_0^z}}=t_{\mathrm{Bloch}}$. 

For $t\gtrsim \frac{|\log(h_\perp)|}{h_\parallel \braket{\sigma^z_0}}$, $V(E)$ varies slowly within this region of dominant support and we may treat $V(E)\approx V_{E=0}$
 as constant inside the integral, yielding
\begin{equation*}
\begin{split}
        \sum_{l} |c_l(t)|^2 &\approx \frac{2}{9} h_{\parallel} \Z \exp\left(-\frac{q}{h_{\parallel} \Z}\right) \int \frac{\sin^2\left(\frac{Et}{2}\right)}{E^2}\,\d E \\
        &\approx \frac{1}{9}  h_{\parallel} \Z t \exp\left(-\frac{q}{h_{\parallel} \Z}\right)  \int_{-\infty}^\infty \frac{\sin^2 x}{x^2}\, \d x  \\
        &= \frac{\pi}{9}  h_{\parallel} \Z t \exp\left(-\frac{q}{h_{\parallel} \Z}\right) \,.
\end{split}
\end{equation*}
We therefore find that for times $t$ obeying
\begin{equation}
    \frac{|\log(h_\perp)|}{h_\parallel \braket{\sigma^z_0}} = t_{\mathrm{lin}} \lesssim t \lesssim t_{\mathrm{Bloch}} = \frac{\pi}{2 h_{\parallel}\braket{\sigma_0^z}} \,,
\end{equation}
the decay rate per volume is approximately constant with value
\begin{equation}  \label{eq:anal_decay_rate}
\gamma_0\approx\frac{\pi}{9}  h_{\parallel} \Z \exp\left(-\frac{|f(i\log(h_\perp))|}{h_\parallel\Z} \right)\,, 
\end{equation}
which is exactly equivalent to result Eq.~(20) in \cite{Rutkevich}.

\begin{figure}
    \centering
    \includegraphics[width=\linewidth]{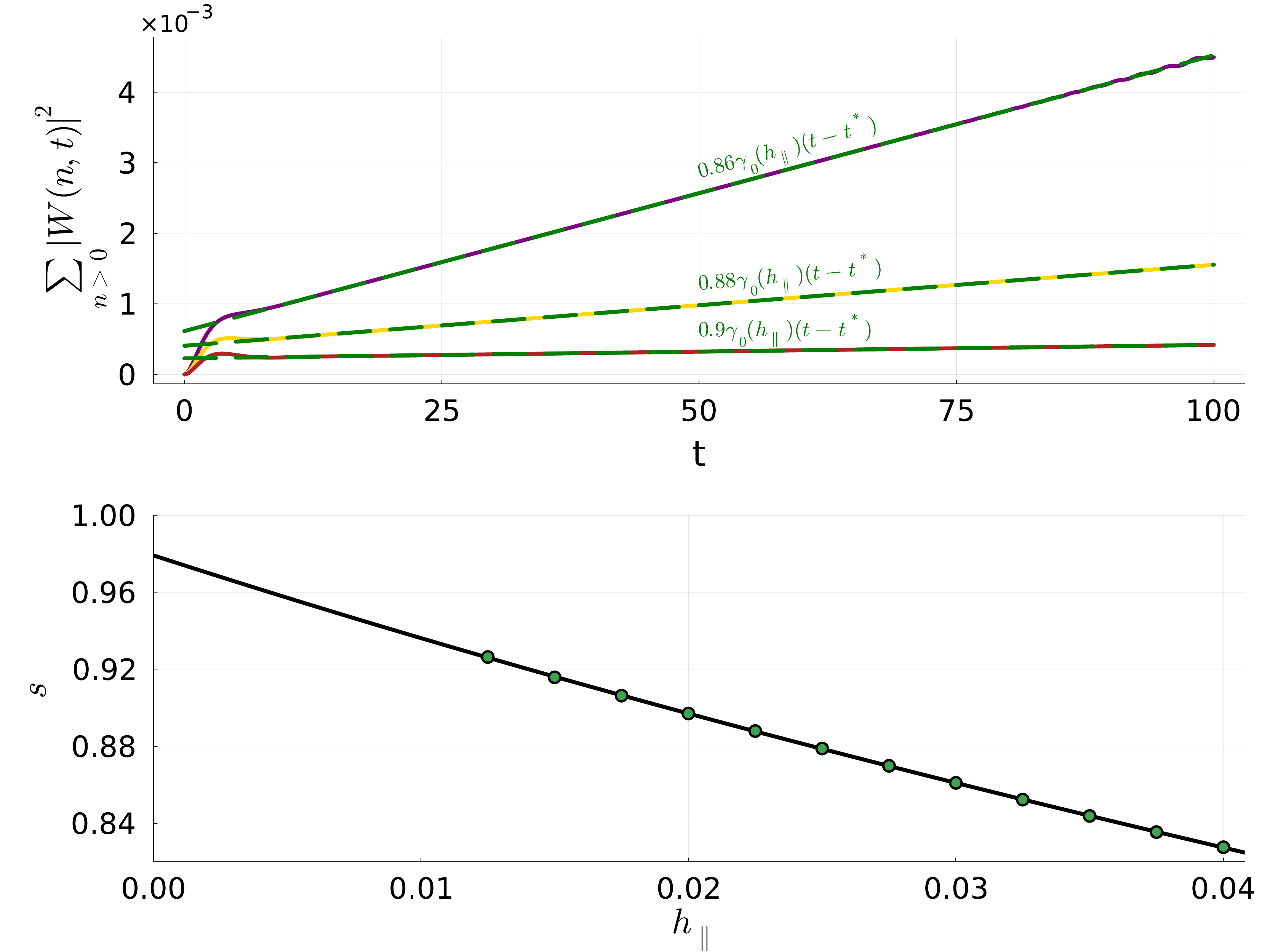}
    \caption{Top panel: total bubble probability $\sum_{n>0} \vert W(n,t) \vert ^2 $ as calculated from the TDVP solution without $F$ and $G$ terms, for fixed $h_\perp=0.8$ and different longitudinal couplings: $h_\parallel=0.03$ (purple), $h_\parallel=0.025$ (gold) and $h_\parallel=0.02$ (red). The best-fit straight lines for $t>1.75t_{\mathrm{lin}}$ (dashed green line) are parametrized as $\gamma(t) = s \gamma_0(h_\parallel) (t-t^*)$ with $s$ and $t^*$ fit parameters and $\gamma_0(h_\parallel)$ given by \eqref{eq:anal_decay_rate}. Bottom panel: extrapolation of the fitted $s$ values for $h_\perp=0.8$ and various $h_\parallel$. Fitting a third order polynomial in $h_\parallel$, we find that $s(h_\parallel=0) \approx 0.98$, close to the theoretical prediction $s=1$ of Eq.~\ref{eq:anal_decay_rate}.}
    \label{fig:prefacfit}
\end{figure}

A comparison with the numerical solution of the linearized TDVP equation is presented in figure \ref{fig:prefacfit} for $h_\perp = 0.8$ and various $h_\parallel$. We find that at times $t \gtrsim t_{\mathrm{lin}}$, the total bubble probability $\sum_{n>0} \vert W(n,t)\vert^2$ becomes linear with a slope that approaches the value predicted by Eq.~\ref{eq:anal_decay_rate} for $h_\parallel\rightarrow 0$. 
    
\section{Summary and conclusions} \label{sec:Conclusions}

To summarize, in this paper we have provided a real space-time window on quantum false vacuum decay for the $d=1+1$ lattice quantum Ising model in terms of true vacuum bubbles. The resulting picture shows rich non-equilibrium dynamics with three qualitatively different stages, roughly delineated by the two time scales $t_{\mathrm{lin}}=|\ln(h_\perp)|/(h_\parallel \Z)$ and $t_{\mathrm{Bloch}}=\pi/(2 h_\parallel \Z)$ (where we only consider large enough values of the transverse field, $e^{-\pi/2}\lesssim h_\perp<1$, such that $t_{\mathrm{lin}}\lesssim t_{\mathrm{Bloch}}$). The initial nucleation ($t\lesssim t_{\mathrm{lin}}$) of small bubbles is followed by a second stage ($t_{\mathrm{lin}}\lesssim t\lesssim t_{\mathrm{Bloch}}$), characterized by a bimodal bubble distribution. In particular, during this stage the distribution shows a static peak at small bubble lengths and a second dynamical peak at larger bubble lengths, with the evolution of the latter signaling semi-classical bubble growth. Furthermore, for certain parameter configurations, we find a non-vanishing tail between the two peaks. As we have shown, this tail can be associated to the production of resonant bubble states, leading to a constant false vacuum decay rate, in quantitative agreement with the result of Rutkevich \cite{Rutkevich} in the limit $h_\parallel \rightarrow 0$. The latter result was originally derived from a computation of the imaginary part of the analytically continued energy of the false vacuum state. Here, we have shown that the very same result arises from a Gaussian approximation of the dynamical evolution of the full quantum state. Finally, during the third stage ($t\gtrsim t_{\mathrm{Bloch}}$), the bubble growth stops due to the lattice phenomenon of Bloch oscillations, leading to a cut-off of the bubble distribution at a maximal bubble length.

In the continuum limit $h_\perp \to 1$ and $h_\parallel \to 0$, we have $\lvert\log h_x \rvert\approx \lvert 1 - h_x \rvert = m$, with $m$ the kink mass (in lattice units), so that we find $t_{\text{lin}} \to m/\sigma$ with $\sigma = h_\parallel \langle \sigma^z \rangle = h_\parallel (1-h_\perp^2)^{1/8}$ the difference in energy density between true and false vacuum. As such, the time scale $t_{\text{lin}} \propto m/\sigma$ has a well defined continuum limit. In contrast, $t_{\text{Bloch}} \to \infty$, in correspondence with the fact that Bloch oscillations are inherent to the lattice. One enticing prospect is to investigate a similar protocol for false vacuum decay in non-relativistic field theories, for example assisted by continuous MPS simulations, to probe the universality of the time scale $t_{\text{lin}} \propto m/\sigma$ in different models.

Our insights can also help to shape future quantum simulation experiments targetting false vacuum decay. Notice however, that in the Gaussian regime that we have considered here, the bubble probabilities are very small, and the observation of bubbles therefore requires large sample sizes. Quenches with larger energy differences between the false and true vacua — and consequently higher bubble production rates — lead to an evolution that cannot be accurately described by the Gaussian ansatz (see e.g.\ Fig.~\ref{fig:GaussianDeviations}), with bubble collisions likely playing a significant role. Further investigation of the non-equilibrium dynamics in this strongly interacting regime is therefore highly desirable.

We finally note that an independent study \cite{Johansen2025} with complementary results appeared on the same day as this submission.

\begin{acknowledgments}
We thank Calvin Chi Chen, Gertian Roose and Iacopo Carussoto for inspiring discussions. Funding for this work has been provided by the Research Foundation Flanders (FWO) [Grant No.\ GOE1520N] and the European Union’s Horizon 2020 program [Grant Agreement No.\ 101125822 (GAMATEN)]. 
\end{acknowledgments}
\bibliography{bib}

\pagebreak
\begin{appendix}
\onecolumngrid
\section{Diagonalization of TFIM} \label{app:sec:Diagonalization}
The Hamiltonian is given by
\begin{equation}
    \hat{H}_0 = -\sum_{n\in\bbZ} \sigma^z_{n} \sigma^z_{n+1} - h_\perp \sum_{n\in\bbZ}  \sigma^x_n\,,
\end{equation}
and can be turned into a quadratic Hamiltonian in fermionic operators by means of a Jordan-Wigner transformation. Before proceeding, it is convenient to first do a rotation in spin space which transforms $\sigma^x \to \sigma^z$ and $\sigma^z \to - \sigma^x$. Using the new notations $\tilde{\sigma}^x = -\sigma^z$ and $\tilde{\sigma}^z = \sigma^x$ to avoid confusion, we arrive at
\begin{equation} \label{app:eq:rotatedH}
    \hat{H}_0 = -\sum_{n\in\bbZ} \tilde{\sigma}^x_{n} \tilde{\sigma}^x_{n+1} - h_\perp \sum_{n\in\bbZ}  \tilde{\sigma}^z_n\,.
\end{equation}
Next we introduce new operators $\tilde{\sigma}^\pm = (\tilde{\sigma}^x\pm i\tilde{\sigma}^y)/2$ which are mapped to the fermionic operators $\hat{c}_n,\hat{c}^\dagger_n$ via the Jordan-Wigner transformation
\begin{equation}
\begin{split}
    \hat{c}_n^\dagger &= \left(\prod_{i<n} \tilde{\sigma}^z_i \right)\tilde{\sigma}^+_n \,, \\
     \hat{c}_n &= \left(\prod_{i<n} \tilde{\sigma}^z_i \right)\tilde{\sigma}^-_n\,,
\end{split}
\end{equation}
with $\hat{n}_j = \hat{c}^\dagger_j \hat{c}_j = (1 + \tilde{\sigma}^z_j)/2$. Plugging this into \eqref{app:eq:rotatedH} and using that $\tilde{\sigma}^z_n =  \hat{c}^\dagger_n \hat{c}_n - \hat{c}_n \hat{c}^\dagger_n$, yields the promised quadratic Hamiltonian
\begin{equation}
    \hat{H}_0 = \sum_{n\in\bbZ} (\hat{c}_n^\dagger - \hat{c}_n)(\hat{c}_{n+1}^\dagger+\hat{c}_{n+1}) + h_\perp \sum_{n\in\bbZ}  (\hat{c}_n \hat{c}^\dagger_n - \hat{c}^\dagger_n \hat{c}_n)\,.
\end{equation}
This Hamiltonian can be brought into a diagonal form with a Fourier transformation, followed by a Bogoliubov transformation
\begin{align}
    \begin{pmatrix}
        \hat{c}_n\\
        \hat{c}_n^\dagger\\
    \end{pmatrix} &= \int_{-\pi}^{\pi} \dpi{k} e^{ikn} 
     \begin{pmatrix}
        \hat{c}(k) \\ \hat{c}^\dagger(-k)
    \end{pmatrix} \\ &= \int_{-\pi}^{\pi} \dpi{k} e^{ikn} 
    U_{\theta_k}
    \begin{pmatrix}
        \hat{\gamma}(k) \\ \hat{\gamma}^\dagger(-k)
    \end{pmatrix} \label{app:eq:bogoliubov}\,,
\end{align}
where
\begin{equation}
     U_{\theta_k} = \begin{pmatrix}
        \cos \theta_k &-i\sin \theta_k \\
        -i\sin \theta_k& \cos \theta_k\\
    \end{pmatrix}\,,
\end{equation}
and then reads
\begin{equation}\label{app:eq:H0}
    \hat{H}_0 = E_0  + \int_{-\pi}^{\pi} \dpi{k}  \omega(k) \hat{\gamma}^\dagger(k)\hat{\gamma}(k)\,,
\end{equation}
with dispersion relation 
\begin{equation} \label{app:eq:disp}
    \omega(k) = 2\sqrt{(1-h_\perp)^2+4h_\perp \sin^2 \frac{k}{2}} = 2\sqrt{(h_\perp-\cos k)^2+\sin^2 k}\,,
\end{equation}
and ground state energy density
\begin{equation}
    e_0 = \frac{E_0}{L} = - \int_0^\pi \dpi{k} \omega(k)\,.
\end{equation}
As usual, the ground state is defined by $\hat{\gamma}(k) | 0_\pm \rangle = 0$ $\forall k$. The Bogoliubov angle $\theta_k$ is uniquely determined by
\begin{equation}
    \begin{split}
        \sin 2\theta_k &= \frac{2\sin k}{\omega(k)}\,, \\
        \cos 2\theta_k &= \frac{2(\cos k-h_\perp)}{\omega(k)}\,, \\
    \end{split}
\end{equation}
and satisfies
\begin{equation}
    \tan 2\theta_k = \frac{\sin k}{\cos k - h_\perp}\,.
\end{equation}
Note that our convention for the Fourier transforms gives
\begin{equation}
\{\hat{\gamma}^\dagger(k), \hat{\gamma}(k') \} = 2\pi \delta(k-k')\,.
\end{equation}
\section{TDVP equation for the squeezed state ansatz} \label{app:sec:TDVP}
The TDVP equation for an unnormalized state $\vert \phi \rangle$ can be found by minimizing the cost function
\begin{equation}
   \left\vert \left\vert \left(i\frac{\partial}{\partial t} - H + \frac{\langle \phi \vert i\frac{\partial}{\partial t} - H \vert \phi \rangle}{\langle \phi \vert \phi \rangle}\right) \vert \phi \rangle \right\vert \right\vert ^2\,.
\end{equation}
Denoting the tangent space by $\{\vert \partial_i \psi \rangle \}$, where $i$ is an abstract index labeling the variational parameters, we can parametrize the time derivative as
\begin{equation}
    \frac{\partial}{\partial t} = \sum_i \Dot{a_i} \vert  \partial_i \phi \rangle\,.
\end{equation}
Working out the cost function in terms of the $\Dot{a_i}$'s and minimizing it by setting its derivative (w.r.t $\Dot{\overline{a_i}}$) to zero, we find that
\begin{equation}
    0 = \sum_j \Dot{a_j} \left( \frac{\langle \partial_i \phi \vert \partial_j \phi \rangle}{\langle \phi \vert \phi \rangle} - \frac{\langle \partial_i \phi \vert \phi \rangle \langle \phi \vert \partial_j \phi \rangle}{\langle \phi \vert \phi \rangle^2}\right) + i
     \left( \frac{\langle \partial_i \phi \vert H \vert \phi \rangle}{\langle \phi \vert \phi \rangle} - \frac{\langle \partial_i \phi \vert \phi \rangle \langle \phi \vert H \vert \phi \rangle}{\langle \phi \vert \phi \rangle^2}\right)\,.
\end{equation}
In our case, the (unnormalized) wave function is given by
\begin{equation}
    \vert \phi \rangle = \exp\left(\int_0^\pi \dpi{k} K(k,t) \gamma^\dagger(k) \gamma^\dagger(-k) \right) \vert 0_\shortminus \rangle \,,
\end{equation}
so that we can identify $a_i \equiv K(k,t)$, $\vert \partial_j \psi \rangle \equiv \gamma^\dagger(k) \gamma^\dagger(-k) \vert \psi \rangle$ and the formal sum $\sum_j$ should be replaced by the integral $\int_0^\pi \dpi{k}$. The TDVP equation then becomes
\begin{multline}
0 = \int_0^\pi \dpi{k} \Dot{K}(k,t) [ \langle \psi \vert \gamma(-q)\gamma(q) \gamma^\dagger(k) \gamma^\dagger(-k) \vert \psi \rangle - \langle \psi \vert \gamma(-q) \gamma(q) \vert \psi \rangle \langle \psi \vert \gamma^\dagger(k) \gamma^\dagger(-k) \vert \psi \rangle] \\+ i [\langle \psi \vert \gamma(-q) \gamma(q) H \vert \psi \rangle - \langle \psi \vert \gamma(-q) \gamma(q) \psi \rangle \langle \psi \vert H \vert \psi \rangle]\,,
\end{multline}
where we introduced the normalized wave function
\begin{equation} \label{wavefunc}
    \vert \psi \rangle = \frac{1}{\sqrt{N}} \exp\left(\int_0^\pi \dpi{k} K(k,t) \gamma^\dagger(k) \gamma^\dagger(-k) \right) \vert 0_\shortminus \rangle\,,
\end{equation}
with normalization
\begin{equation}
    N = \prod_{k>0} (1+\vert K(k,t) \vert^2) = \exp \left(L \int_0^\pi \dpi{k} \log[1+\vert K(k,t)\vert^2] \right) \,,
\end{equation}
where $L$ is the formally infinite length of the system. Defining the connected parts,
\begin{align}
      \langle \psi \vert \gamma(-q)\gamma(q) \gamma^\dagger(k) \gamma^\dagger(-k) \vert \psi \rangle_c &= \langle \psi \vert \gamma(-q)\gamma(q) \gamma^\dagger(k) \gamma^\dagger(-k) \vert \psi \rangle - \langle \psi \vert \gamma(-q)\gamma(q) \vert \psi \rangle \langle \psi \vert \gamma^\dagger(k) \gamma^\dagger(-k) \vert \psi \rangle\,, \\
       \langle \psi \vert \gamma(-q) \gamma(q) H \vert \psi \rangle_c &=  \langle \psi \vert \gamma(-q) \gamma(q) H \vert \psi \rangle -  \langle \psi \vert \gamma(-q) \gamma(q) \vert \psi \rangle \langle \psi \vert H \vert \psi \rangle\,,
\end{align}
we find after a straightforward calculation that
\begin{equation}
  \langle \psi \vert \gamma(-q)\gamma(q) \gamma^\dagger(k) \gamma^\dagger(-k) \vert \psi \rangle_c = 2\pi \delta(0)\frac{2\pi \delta(k-q)-2\pi\delta(k+q)}{(1+\vert K(k,t) \vert ^2)^2}\,.
\end{equation}
For the other connected term, we use that the Hamiltonian can be brought into a partially diagonal form,
\begin{equation}
    H = H_0 - h_\parallel \sum_n \sigma_n^z\,,
\end{equation}
with $H_0$ from equation \eqref{app:eq:H0}. Then
\begin{equation}
    \langle \psi \vert \gamma(-q) \gamma(q) H \vert \psi \rangle_c =  \langle \psi \vert \gamma(-q) \gamma(q) H_0 \vert \psi \rangle_c - 2\pi \delta(0) h_\parallel \langle \psi \vert\gamma(q) \gamma(-q) \sigma^z_0\vert \psi \rangle_c\,,
\end{equation}
where it can be worked out that
\begin{equation}
    \langle \psi \vert \gamma(-q) \gamma(q) H_0 \vert \psi \rangle_c = 2\pi \delta(0)\frac{2K(q,t)\omega(q)}{(1+\vert K(k,t)\vert^2)^2}\,.
\end{equation}
After collecting these results, the TDVP equation for the squeezed state reduces to
\begin{equation} \label{app:eq:TDVP}
    0 = \frac{\Dot{K}(q,t)}{(1+\vert K_q \vert^2)^2} + i\left(\frac{2K(q,t)\omega(q)}{(1+\vert K_q \vert^2)^2}  +h_\parallel  \langle \psi\vert\gamma(q) \gamma(-q) \sigma^z_0\vert \psi \rangle_c\right)\,.
\end{equation}
Here, the last term cannot be worked out directly due to the fact that in the fermionic language, the $\sigma^z_0$ operator contains a half infinite Jordan-Wigner string. However, it can be shown that  $\sigma^z_0$ can be written as a normal ordered exponential of the fermionic ladder operators \cite{Jimbo}. Hence, it is possible to calculate the desired matrix element using fermionic coherent states and Grasmann integrals, giving a closed form expression in terms of pfaffians. These can then be expanded up to the desired order in $K(k,t)$, allowing the TDVP equation to be solved to any order. This strategy is particularly useful (though involved) for calculating up to second order and higher. For the first order expansion of \eqref{app:eq:TDVP}, it is easier to insert the corresponding order of the squeezed state and use the known expressions for the form factors. This is done in the next section.

\section{Derivation of the linearized TDVP equation} \label{app:sec:TDVP_solution}
As discussed at the end of the previous section, we will use the first order expansion of the squeezed state,
\begin{equation} \label{app:eq:1order_state}
    \vert \psi(t) \rangle \approx \vert 0_\shortminus \rangle + \int_0^\pi \dpi{k} K(k,t) \vert k,-k\rangle\,,
\end{equation}
to work out the matrix element $\langle \psi(t)\vert\gamma(q) \gamma(-q) \sigma^z_0\vert \psi(t) \rangle_c$. We obtain
\begin{equation} \label{app:eq:first_order_zterm}
    \langle \psi(t) \vert \gamma(q) \gamma(-q) \sigma^z_0 \vert \psi(t)\rangle \approx \Zr{q,-q} + \int_0^\pi \dpi{k} K(k,t) \Zm{q,-q}{k,-k} + \int_0^\pi \dpi{k}\overline{K(k,t)} \Zr{-k,k,q,-q} \,,
\end{equation}
and
\begin{equation}
     \langle \psi(t) \vert \gamma(q) \gamma(-q) \vert \psi(t) \rangle \langle \psi(t) \vert \sigma^z_0 \vert \psi(t)\rangle =  -2\pi \delta(0) \frac{K(q,t)}{1+\vert K(q,t)\vert^2} \langle \psi(t) \vert \sigma^z_0 \vert \psi(t)\rangle \approx 2\pi \delta(0) K(q,t) \Z \,.
\end{equation}
The two-point functions take the form
\begin{align} \label{app:eq:div_matrix_el}
   \Zl{q,-q} &=-\Z F(\vert q,-q)\,, \\
   \Zr{q,-q} &=-\Z F(q,-q\vert)\,, \\
     \Zm{q}{k} &=-\Z F(q\vert k)\,,
\end{align}
with $\langle 0_\shortminus \vert \sigma^z_0 \vert 0_\shortminus \rangle = -\Z = -(1-h_\perp^2)^{1/8}$. The functions $F(\vert k,k')$, $F(k,k' \vert)$ and $F(k\vert k')$ are given by\footnote{Note that these form factors, taken from \cite{Calabrese_2012} and references therein, differ from the ones presented in \cite{Rutkevich2008,Bastianello_squeezed}. The apparent discrepancy is due to different choices for the basis of free fermion states, related by the unitary transformation $\gamma(k)\to ie^{-ik/2}\gamma(k)$.}
\begin{align}
    \label{Fl}
    F(\vert k, k') &= \frac{\omega(k)-\omega(k')}{\sqrt{\omega(k)\omega(k')}} \frac{e^{-i(k+k')/2}}{1-e^{-i(k+k')}} = \frac{-8ih_\perp \sin(\frac{k-k'}{2})}{(\omega(k)+\omega(k'))\sqrt{\omega(k)\omega(k')}}\,,\\
    \label{Fr}
    F(k, k' \vert ) &= \frac{\omega(k)-\omega(k')}{\sqrt{\omega(k)\omega(k')}} \frac{e^{i(k+k')/2}}{e^{i(k+k')}-1} = - \overline{ F(\vert k, k')}\,,\\ \label{Fm}
    F(k\vert k') &= \frac{\omega(k)+\omega(k')}{\sqrt{\omega(k)\omega(k')}}  \frac{e^{i(k-k')/2}}{1-e^{i(k-k')}} =  \frac{\omega(k)+\omega(k')}{\sqrt{\omega(k)\omega(k')}} \frac{i}{2\sin(\frac{k-k'}{2})}\,.
\end{align}
Using the afore mentioned fermionic coherent states, or equivalently using Wick's theorem, it can be shown that
\begin{align} \label{app:eq:div_matrix_el2}
   \Zm{q,-q}{k,-k} &= -\Z F(q,-q\vert k,-k)\,, \\
     \Zm{-k,k}{q,-q} &= -\Z F(-k,k,q,-q\vert )\,,
\end{align}
where we introduced
\begin{align} \label{app:eq:div_FF}
   F(q,-q\vert k,-k) = F(-q\vert k) F(k\vert -q) - F(q\vert k) F(-q\vert-k) + F(q,-q\vert)F(\vert k,-k)\,,
\end{align}
and
\begin{equation}
    F(-k,k,q,-q\vert ) = F(k,q\vert) F(-k,-q\vert ) -F(k,-q\vert)F(-k,q\vert) + F(q,-q\vert)F(-k,k\vert)\,.
\end{equation}
The first order approximation in $K(k,t)$ of \eqref{app:eq:TDVP} is then given by
\begin{multline} \label{app:eq:first_order_TDVP}
    i\Dot{K}(q,t) =2K(q,t)\omega(q) - h_\parallel\Z \biggl(F(q,-q\vert)+\int_0^\pi \dpi{k} K(k,t) F(q,-q\vert k, -k) \\+ \int_0^\pi \dpi{k}\overline{K(k,t)}F(-k,k,q,-q\vert) + 2\pi \delta(0)K(q,t)\biggr) \,.
\end{multline}
Note that $F(k\vert k')$ is singular when $k=k'$ and requires special treatment. To understand this divergence, it is instructive to consider the matrix element before the thermodynamic limit was taken i.e.\ for the TFIM on a finite, periodic chain with length $L$. There, the Fock space splits into two subspaces with different fermion parity. The sector with even parity (even number of fermions) is called the Neveu-Schwarz (NS) sector. Its basis can be labeled with momenta $k$, given by
\begin{equation}
    q_j = \frac{\pi}{L} (2j+1), \quad j=-\frac{L}{2},\dots,\frac{L}{2}-1\,.
\end{equation}
The other sector has uneven parity (uneven number of fermions) and is referred to as the Raymond sector (R). The momenta $k\in$ R are quantized according to
\begin{equation}
    k_m = \frac{2\pi m}{L} , \quad m=-\frac{L}{2},\dots,\frac{L}{2}-1\,.
\end{equation}
Since the $\sigma^z$ operator flips the fermion parity, the only non-vanishing matrix elements are of the form
\begin{equation}
    \prescript{}{\text{NS}}{\langle} q_1,\dots ,q_N \vert \sigma_0^z \vert k_1,\dots ,k_M \rangle_\text{R}\,,
\end{equation}
and can be found in \cite{Calabrese_2012}. For further details on the diagonalization of the finite chain  TFIM, we refer the reader to appendix A of the aforementioned reference. Consider now the part of 
\eqref{app:eq:first_order_TDVP} that diverges,
\begin{equation}
    \int_0^\pi \dpi{k} K(k,t) \left[F(q\vert-k)F(-q\vert k)- F(q\vert k)F(-q\vert -k)\right]= \sum_{n\in \bbZ} W(n,t)  S(n)\,,
\end{equation}
where
\begin{equation}
    S(n) = \int_{-\pi}^\pi \dpi{k} e^{-ikn} F(q\vert-k)F(-q\vert k)\,,
\end{equation}
and we used that $K(-k,t)=-K(k,t)$. Its finite size counter part is given by
\begin{equation}
    S_L(n) = \frac{1}{L}\sum_{k\in \text{R}} e^{-ikn} F(q\vert-k)F(-q\vert k) =  \frac{1}{L} \sum_{k\in \text{R}}  \frac{(\omega(q)+\omega(k))^2}{\omega(q)\omega(k)} \frac{e^{-ikn}}{4\sin^2(\frac{q+k}{2})}\,.
\end{equation}
The problematic part of the sum can be singled out via the subtraction procedure
\begin{equation} \label{app:eq:SL_substracted}
     S_L(n) =  \frac{1}{L} \sum_{k\in \text{R}} \left[ \frac{(\omega(q)+\omega(k))^2}{\omega(q)\omega(k)} -4\right]\frac{e^{-ikn}}{4\sin^2(\frac{q+k}{2})} + \frac{4}{L}  \sum_{k\in \text{R}} \frac{e^{-ikn}}{4\sin^2(\frac{q+k}{2})}\,.
\end{equation}
Since
\begin{equation}
    \frac{(\omega(q)+\omega(k))^2}{\omega(q)\omega(k)} = 4 + \left(\frac{\omega'(k)}{\omega(k)}\right)^2(q+k)^2 + \mathcal{O}{(q+k)^3}\,,
\end{equation}
the first term in \eqref{app:eq:SL_substracted} is finite in thermodynamic limit. The second term can be calculated analytically in the $L\to \infty$ limit by means of complex contour integration. For this, we use the function
\begin{equation}
    f(z) = \frac{1}{2\pi} \frac{e^{-izn}}{4\sin^2(\frac{q+z}{2})} \frac{1}{1-e^{-\text{sign}(n)iLz}}\,,
\end{equation}
and take a box contour with corners $-\pi+iR,\pi+iR,\pi-iR$ and $-\pi-iR$ and let $R\to\infty$. We find 
\begin{equation}
    \frac{1}{L}  \sum_{k\in \text{R}} \frac{e^{-ikn}}{4\sin^2(\frac{q+k}{2})} = -\frac{\vert n \vert}{2}e^{iqn} + \frac{L}{4}e^{iqn}\,,
\end{equation}
from which it follows that
\begin{equation} 
   S(n) = \lim_{L\to\infty}  S_L(n) = \int_{-\pi}^\pi \dpi{k} \left[ \frac{(\omega(q)+\omega(k))^2}{\omega(q)\omega(k)} -4\right]\frac{e^{-ikn}}{4\sin^2(\frac{q+k}{2})} - 2\vert n \vert e^{iqn} + 2\pi \delta(0)  e^{iqn}\,.
\end{equation}
The correct interpretation of the diverging term in \eqref{app:eq:first_order_TDVP} is therefore
\begin{equation} \label{eq:Freg}
    \int_0^\pi \dpi{k} K(k,t) F(q,-q\vert k,-k) = -2\pi \delta(0) K(q,t) + \int_{0}^\pi \dpi{k}K(k,t)  F(q,-q\vert k,-k)_\text{reg}\,,
\end{equation}
with
\begin{equation}  
    \int_{0}^\pi \dpi{k}K(k,t) F(q,-q\vert k,-k)_\text{reg} = 
    \sum_{n\in\mathbb{Z}} 2\vert n \vert W(n,t) e^{-iqn} + \int_{0}^\pi \dpi{k}K(k,t)  G(q,k)\,,
\end{equation}
where we used that $W(-n,t)=-W(n,t)$ and defined the kernel
\begin{equation}
    G(q,k) = \left[ \frac{(\omega(q)+\omega(k))^2}{\omega(q)\omega(k)}-4\right] \left[\frac{1}{{4\sin^2(\frac{q+k}{2})}} - \frac{1}{{4\sin^2(\frac{q-k}{2})}}\right] + F(q,-q\vert)F(\vert k,-k)\,.
\end{equation}
Here, $G(q,k)$ represents quantum corrections to the naive interpretation of a bubble as two particles with momenta $q$ and $-q$, subject to the semi-classical Hamiltonian
\begin{equation}
    H = 2\omega(q) - 2h_\parallel \Z  \vert x \vert\,,
\end{equation}
where $x$ is the distance between the two particles. Putting everything together, we find that the linearised TDVP equation takes the form
\begin{multline} \label{app:eq:1order_TDVP_num}
    i\Dot{K}(q,t) =2K(q,t)\omega(q) - 2h_\parallel\Z \sum_{n\in\mathbb{Z}} \vert n \vert W(n,t) e^{-iqn} - h_\parallel \Z F(q,-q\vert)  \\- h_\parallel\Z \biggl(\int_0^\pi \dpi{k} K(k,t) G(q,k) + \int_0^\pi \dpi{k}\overline{K(k,t)}F(-k,k,q,-q\vert)\biggr) \,.
\end{multline}
This integro-differential equation can be solved numerically by a combination of one's favorite ODE solver and discretization scheme for $q$ and $k$. We used the toolbox of \textit{ApproxFun.jl} \cite{ApproxFun} to implement the different operators in \eqref{app:eq:1order_TDVP_num} and performed the timestepping with the help of \textit{DifferentialEquations.jl} \cite{Differentialequations}. In real-space, the linearized TDVP equation is given by
\begin{equation}  \label{app:eq:1order_TDVP_realspace}
\begin{split}
        i\Dot{W}(n,t) =2 \sum_{n'\in\mathbb{Z}}T_{n-n'}W(n',t) - 2h_\parallel\Z \vert n \vert &W(n,t) -  \frac{1}{2}  h_\parallel \Z\text{sign}(n)h_\perp^{\vert n \vert} \\ &- h_\parallel\Z \sum_{n,n'\in \mathbb{Z}} \biggl( G_{n,n'}W(n',t) + F_{n,n'}\overline{W(n',t)} \biggr)\,,
\end{split}
\end{equation}
with
\begin{equation}
    T_{n-n'} = \int_{-\pi}^\pi \dpi{k} \omega(k) e^{ik(n-n')}\,,
\end{equation}
\begin{equation}
    G_{n,n'} = \frac{1}{2}\int_{-\pi}^\pi \dpi{q} e^{iqn} \int_{-\pi}^\pi \dpi{k} e^{-ikn'} G(q,k)\,,
\end{equation}
and
\begin{equation}
    F_{n,n'} =  -\frac{1}{2}\int_{-\pi}^\pi \dpi{q} e^{iqn} \int_{-\pi}^\pi \dpi{k} e^{-ikn'} F(-k,k,q,-q\vert)\,.
\end{equation}
In deriving \eqref{app:eq:1order_TDVP_realspace}, we used that
\begin{equation}
    \int_{-\pi}^\pi \dpi{k} e^{ikn} F(k,-k\vert) = \frac{1}{2} \text{sign}(n)h_\perp^{\vert n \vert}\,,
\end{equation}
as can be established via complex contour integration. 
\\
\\The link to earlier works \cite{Rutkevich,Bastianello_squeezed,Wilczek} can be made when $h_\perp$ and $h_\parallel$ are small. In that regime, $G_{n,n'}$ and $F_{n,n'}$ can safely be neglected and we obtain a regular ODE
\begin{equation} \label{app:eq:first_order_TDVP_ODE}
    i\Dot{W}(n,t) =2 \sum_{n'\in\mathbb{Z}}T_{n-n'}W(n',t) - 2h_\parallel\Z \vert n \vert W(n,t) - \frac{1}{2}  h_\parallel \Z\text{sign}(n)h_\perp^{\vert n \vert}\,,
\end{equation}
If we define the functions $\phi_l(n)$ ($l\in \mathbb{N}$) as the eigenvectors of the eigenvalue problem
\begin{equation} \label{app:eq:eigprob_n}
    E_l\phi_l(n)=\sum_{n'\in\mathbb{Z}}2T_{n-n'}\phi_l(n') - 2 h_\parallel\Z \vert n \vert \phi_l(n)\,,
\end{equation}
with normalization
\begin{equation}
    \sum_{n>0} \phi_l(n) \phi_{l'}(n) = \delta_{l,l'}\,,
\end{equation}
then the solution of \eqref{app:eq:first_order_TDVP_ODE} is given by
\begin{equation} \label{app:eq:Wsol}
    W(n,t) = \sum_l c_l(t)\phi_l(n)\,,
\end{equation}
with
\begin{equation}
 c_l(t) = \left(\frac{e^{-iE_l t}-1}{E_l} \right) V_l\,,
\end{equation}
and
\begin{equation} \label{eq:Vl_eigenv}
    V_l =  -\frac{1}{2}h_\parallel\Z  \sum_{n>0} \phi_l(n) h_\perp^n \,.
\end{equation}
Note that \eqref{app:eq:eigprob_n} is exactly the eigenvalue problem that determines the two-particle spectrum reported in \cite{Rutkevich,Wilczek}. From \eqref{app:eq:Wsol} it follows that
\begin{equation} \label{app:eq:Ksol}
    K(k,t) =  \sum_l c_l(t)\phi_l(k)\,,
\end{equation}
with $\phi_l(k)$ the Fourier transform of $\phi_l(n)$. Finally, plugging \eqref{app:eq:Wsol} into the \eqref{app:eq:1order_state} gives Eq.\ (6) in \cite{Wilczek}
\begin{equation}
    \vert \psi(t) \rangle \approx \vert 0_\shortminus \rangle + \sum_l c_l(t) \vert \phi_l \rangle\,,
\end{equation}
where we set
\begin{equation}
    \vert \phi_l \rangle = \int_{0}^\pi \dpi{k} \phi_l(k) \vert k,-k \rangle\ = \frac{1}{2} \sum_{j,n} \phi_l(n) b^\dagger_{n+j} b^\dagger_{j} \vert 0 \rangle\,.
\end{equation}

\section{Correlation matrix and structure factors} \label{app:sec:structfactor}
The time-dependent structure factor for the operators $\hat{c}^\dagger$ and $\hat{c}$, is given by
\begin{align}
    s^{\shortplus\shortminus}(k,t) &= \frac{1}{\vert \bbZ\vert} \sum_{m,n} e^{-ik(m-n)}\langle \psi(t) \vert c^\dagger_n c_m \vert \psi(t)\rangle \\
    &= \sum_\Delta e^{-ik\Delta} \langle c^\dagger_0 c_\Delta \rangle_t\,,
\end{align}
where in the second line we used translational invariance to eliminate one infinite sum. In momentum space $s^{\shortplus\shortminus}(k,t)$ reduces to
\begin{equation}
    s^{\shortplus\shortminus}(k,t) = \int_{-\pi}^{\pi}\dpi{q} \langle c^\dagger(-q) c(k) \rangle_t \,.
\end{equation}
The two-point function $\langle c^\dagger(-q) c(k) \rangle_t$ can be determined by means of the following correlators
\begin{align} \label{app:eq:correlators_gg}
    \langle\gamma(q)\gamma(k)\rangle_t &= 2\pi \delta(k+q) \frac{K(k,t)}{1+\vert K(k,t)\vert^2}\,, \\ \label{app:eq:correlators_gdg}
    \langle\gamma^\dagger(-q)\gamma(k)\rangle_t &= 2\pi \delta(k+q) \frac{\vert K(k,t) \vert^2}{1+\vert K(k,t)\vert^2}\,, \\ \label{app:eq:correlators_ggd}
    \langle\gamma(q)\gamma^\dagger(-k)\rangle_t &= 2\pi \delta(k+q) \frac{1}{1+\vert K(k,t)\vert^2}\,, \\ \label{app:eq:correlators_gdgd}
    \langle\gamma^\dagger(-q)\gamma^\dagger(-k)\rangle_t &= 2\pi \delta(k+q) \frac{\overline{K(k,t)}}{1+\vert K(k,t)\vert^2}\,,
\end{align}
Using the Bogoliubov transformation in \eqref{app:eq:bogoliubov}, we obtain
\begin{equation} \label{app:eq:spm}
    s^{\shortplus\shortminus}(k,t) = \frac{\cos2\theta_k \vert K(k,t)\vert^2 +\sin 2\theta_k \text{Im}(K(k,t))}{1+\vert K(k,t)\vert^2}+ s^{\shortplus\shortminus}(k,0) \,,
\end{equation}
with $s^{\shortplus\shortminus}(k,0)=\sin^2 \theta_k$ the structure factor at $t=0$ and $\text{Im}(K(k,t))$ the imaginary part of $K(k,t)$. More generally, the Fourier transform
of the time-dependent correlation matrix 
\begin{equation}
    \Gamma_{n,m}(t) = \begin{pmatrix}
         \langle \hat{c}_n\hat{c}^\dagger_m\rangle & \langle \hat{c}_n\hat{c}_m \rangle \\ \langle \hat{c}^\dagger_n\hat{c}^\dagger_m \rangle & \langle \hat{c}^\dagger_n\hat{c}_m \rangle
    \end{pmatrix}\,,
\end{equation}
can be written as
\begin{align}
    \Gamma(k,t) &= \frac{1}{\vert \bbZ\vert} \sum_{m,n} e^{-ik(m-n)} \Gamma_{n,m}(t) \\
    &= \sum_\Delta e^{-ik\Delta} \Gamma_{0,\Delta}(t) \\ &= \int_{-\pi}^{\pi} \dpi{q}
    \begin{pmatrix}
        \langle \hat{c}(q)\hat{c}^\dagger(-k)\rangle & \langle \hat{c}(q)\hat{c}(k) \rangle \\ \langle \hat{c}^\dagger(-q)\hat{c}^\dagger(-k) \rangle & \langle \hat{c}^\dagger(-q)\hat{c}(k) \rangle
    \end{pmatrix} \\
    &=  \frac{1}{1+\vert K(k,t)\vert^2} U_{\theta_{-k}}
    \begin{pmatrix}
        1 & K(k,t) \\ \overline{K(k,t)} & \vert K(k,t) \vert^2
    \end{pmatrix}
    U_{\theta_{-k}}^\dagger
\end{align}
The elements of $\Gamma(k,t)$, labeled according to
\begin{equation}
\Gamma(k,t) = 
    \begin{pmatrix}
       s^{\shortminus\shortplus}(k,t) &  s^{\shortminus\shortminus}(k,t) \\
        s^{\shortplus\shortplus}(k,t) &  s^{\shortplus\shortminus}(k,t)
    \end{pmatrix} \,,
\end{equation}
can be efficiently calculated using MPS with perfect resolution in the momentum. We have that the wave function $K(k,t)$ can be directly determined from the MPS simulations as
\begin{equation}
    K_{\text{MPS}}(k,t) = \frac{[D]_{1,2}}{[D]_{1,1}}\,,
\end{equation}
where $[D]_{i,j}$ are the elements of the matrix
\begin{equation}
    D = U_{\theta_{-k}}^\dagger \Gamma(k,t) U_{\theta_{-k}}\,.
\end{equation}
In order to numerically reconstruct the real-space wave function from $K_{\text{MPS}}(k,t)$, we consider the natural truncation,
\begin{equation}
    K(k,t) \approx -2i \sum_{n=1}^{N}\sin(kn) W(n,t)\,,
\end{equation}
of the infinite Fourier series for $K(k,t)$, motivated by the fact that the bubble amplitudes we encounter have a finite support, that grows in time until $t_{\text{Bloch}}$. From this it follows that $W_{\text{MPS}}(n,t)$ can be found by a discrete Fourier (sine) transform
\begin{equation} \label{app:eq:DFT}
    W_{\text{MPS}}(n,t) = \frac{i}{N} \sum_{m=0}^{N-1} K_{\text{MPS}}\left(\frac{m\pi}{N} ,t \right) \sin\left(\frac{m\pi}{N} n\right)\,.
\end{equation}

\section{Transversal and longitudinal magnetization} \label{app:sec:magnetization}
 
\begin{figure}[t]
    \centering
    \includegraphics[width=0.65\linewidth]{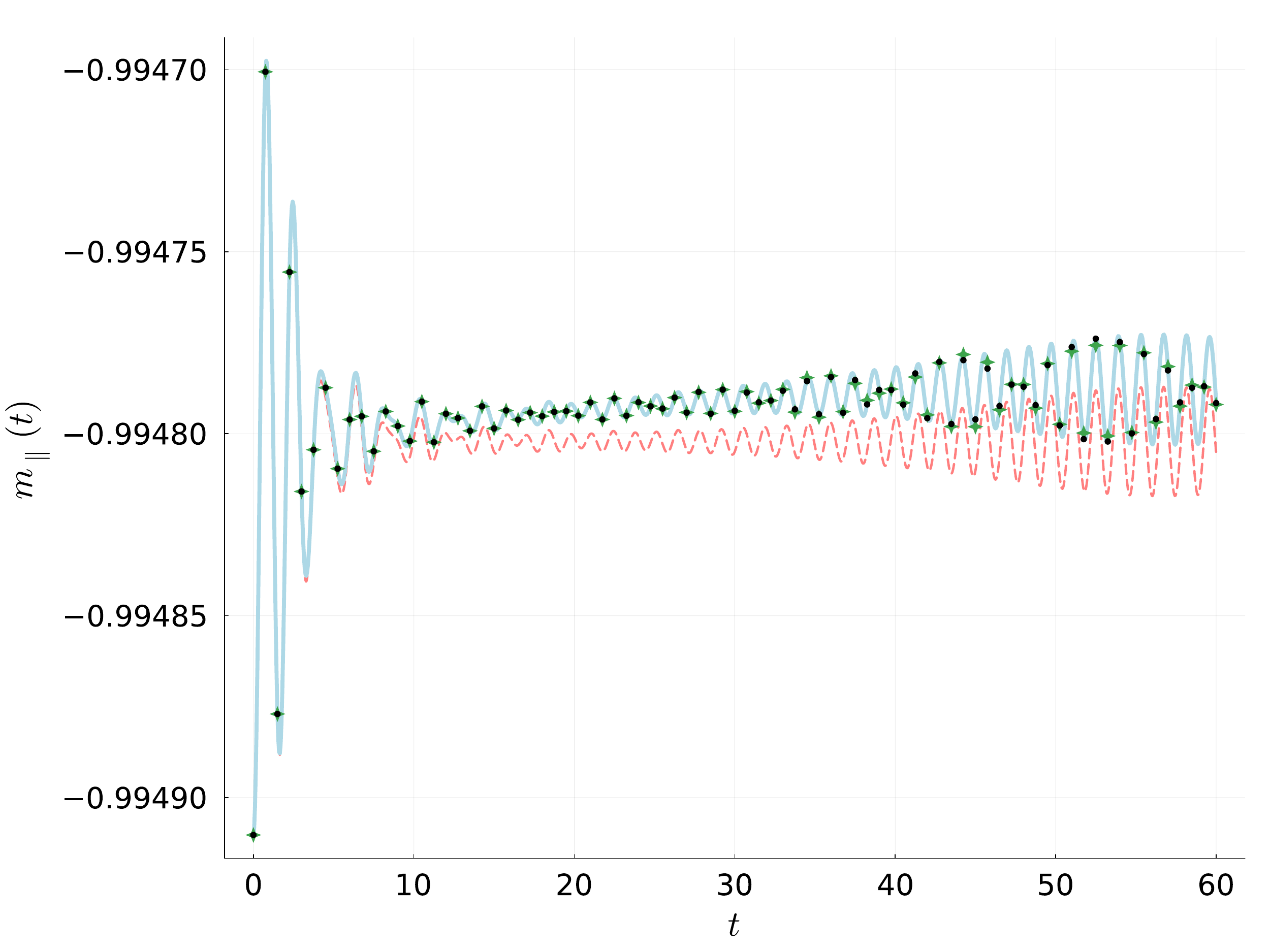}
    \caption{Comparison between \eqref{app:eq:analytical_zmag_final} (blue line) and the longitudinal magnetization obtained from MPS simulation, for parameters $h_\perp=0.2$ and $h_\parallel=0.02$ with Schmidt cut parameters $\eta=10^{-6}$ (green stars) and $\eta=10^{-6}$ (black dots). The first order approximation \eqref{app:eq:Z_1order} is shown with the red dashed line.}
    \label{app:fig:zmag}
\end{figure}

The time-dependent transversal magnetization $m_\perp(t)$,
\begin{equation}
    m_\perp(t)=\frac{1}{\vert \bbZ \vert} \sum_i\langle \psi(t) \vert \sigma^x_i \vert \psi(t) \rangle = \langle \psi(t) \vert \sigma^x_0 \vert \psi(t) \rangle \,,
\end{equation}
can be easily evaluated using (\ref{app:eq:correlators_gg}--\ref{app:eq:correlators_gdgd}) since $\sigma^x=\tilde{\sigma}^z$ is a quadratic fermion operator. We find that
\begin{equation} \label{app:eq:analytical_xmag}
    1+m_\perp(t) =  2\int_{-\pi}^\pi \dpi{k} s^{\shortplus\shortminus}(k,t)\,,
\end{equation}
with $s^{\shortplus\shortminus}(k,t)$ as defined in \eqref{app:eq:spm}. We remark that obtaining the longitudinal magnetization is significantly more challenging due to the fact that, in the fermionic language, $\sigma^z$ consists of an odd number of Majorana operators. Consequently, one must consider correlation functions of the form $\langle \sigma^z_i \sigma^z_{i+\Delta} \rangle $ and determine $m_\parallel(t)=\langle \sigma^z_0(t) \rangle$ as $\lim_{\Delta \to \infty } \sqrt{\langle \sigma^z_i \sigma^z_{i+\Delta} \rangle}$. While, in principle, this could be evaluated using Wick's theorem—reducing the problem to finding the determinant of an infinite block Toeplitz matrix—this approach proves to be too complex in practice. A much simpler approach is to insert the squeezed state ansatz into $m_\parallel(t)=\langle \psi(t) \vert \sigma_0^z \vert \psi(t) \rangle$. Taking,
\begin{align}
    \vert \psi(t) \rangle \approx \vert \psi^{(1)}(t) \rangle &= \frac{1}{\sqrt{N}} \left(\vert 0_\shortminus \rangle + \int_0^\pi \dpi{k} K(k,t) \vert k,-k\rangle \right)  \\ &= \left(1-\pi \delta(0) \int_0^\pi \dpi{k} \vert K(k,t) \vert^2 \right)\vert 0_\shortminus \rangle + \int_0^\pi \dpi{k} K(k,t) \vert k,-k\rangle\,,
\end{align}
we have that the corresponding approximation of the longitudinal magnetization, $ m^{(1)}_\parallel(t)$, reads
\begin{equation} \label{app:eq:Zexpansion}
    m^{(1)}_\parallel(t) = -\Z \left[ \int_0^\pi \dpi{k} \biggl(F(\vert k,-k )K(k,t)+ \text{h.c.}\biggr) + \int_0^\pi \dpi{k} K(k,t) \int_0^\pi \dpi{q} \overline{K(q,t)}  F(-q,q\vert k,-k)_\text{reg} \right]\,,
\end{equation}
where \eqref{eq:Freg} was used to cancel the divergencies. For small $\vert K(k,t) \vert$, the quadratic term can be neglected and \eqref{app:eq:Zexpansion} recovers the result from \cite{Wilczek},
\begin{equation} \label{app:eq:Z_1order}
    m^{(1)}_\parallel(t) \approx -\Z - \biggl(\sum_l c_l(t) V_l + \text{h.c.}\biggr)\,,
\end{equation}
where we used \eqref{app:eq:Ksol}.
The time-derivative of the longitudinal magnetization,
\begin{equation} \label{app:eq:mdot}
    \Dot{m}^{(1)}_\parallel(t) = \Z\int_0^\pi \dpi{q} \overline{\Dot{K}(q,t)} \left[ \int_0^\pi \dpi{k} K(k,t) F(q,-q\vert k,-k)_\text{reg} + F(q,-q\vert)\right] + \text{h.c.}\,,
\end{equation}
can be used to cast \eqref{app:eq:Zexpansion} into a simpler form. After using that $K(k,t)$ satisfies the linearized TDVP equation \eqref{app:eq:1order_TDVP_num}, we obtain
\begin{multline} 
 \Dot{m}^{(1)}_\parallel(t) = \frac{2}{h_\parallel} \int_0^\pi \dpi{k} [K(q,t)\overline{\Dot{K}(q,t)} + \Dot{K}(q,t)\overline{K(q,t)}] \omega(q)
      \\ + \Z \left( \int_0^\pi \dpi{q} \Dot{K}(q,t) \int_0^\pi \dpi{k} K(k,t) F(\vert q,-q,k,-k) +\text{h.c.}\right)\,.
\end{multline}
\begin{figure}[t]
    \centering
    \includegraphics[width=0.775\linewidth]{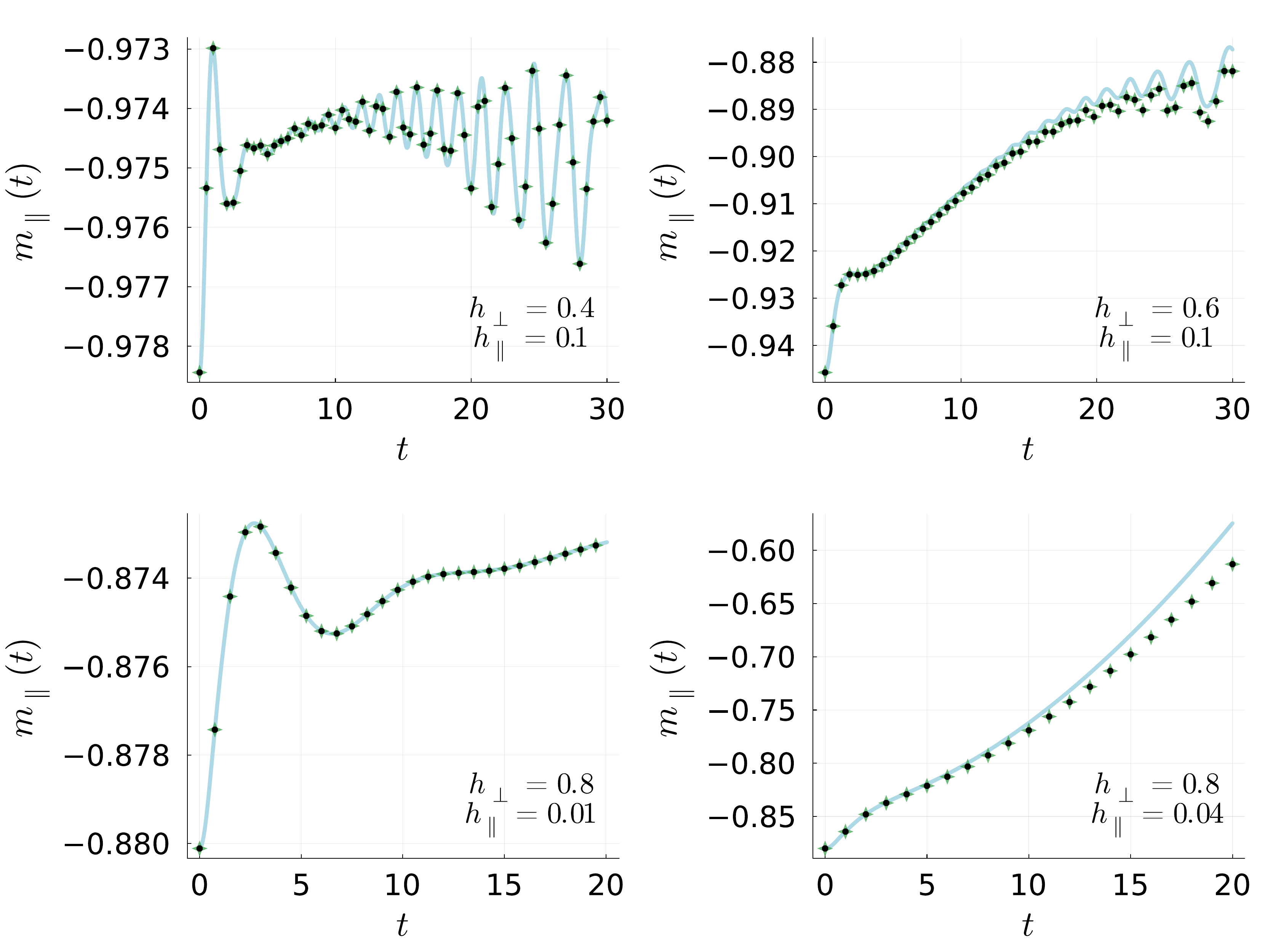}
    \caption{Longitudinal magnetization $m_\parallel(t)$ of the squeezed state ansatz \eqref{app:eq:analytical_zmag_final} (blue line) and MPS data for different $h_\perp$ and $h_\parallel$. MPS simulations were obtained for $\eta = 10^{-5}$ (green stars) and $\eta = 10^{-6}$ (black dots), each time with $D_\mathrm{max}=250$, except for $(h_\perp,h_\parallel)=(0.8,0.04)$ where $D_\mathrm{max}=500$ and the black dots show results for $\eta = 5\times 10^{-6}$.}
    \label{app:fig:many_zmag}
\end{figure}
From 
\begin{equation}
   \frac{\d}{\d{t}} \vert K(q,t)\vert^2 = K(q,t)\overline{\Dot{K}(q,t)}+\Dot{K}(q,t)\overline{K(q,t)}\,,
\end{equation}
and
\begin{equation}
    \frac{\d}{\d{t}} \biggl(K(q,t)K(k,t) \biggr) = \Dot{K}(q,t)K(k,t) + K(q,t)\Dot{K}(k,t)\,,
\end{equation}
it follows that the solution of \eqref{app:eq:mdot} is
\begin{equation} \label{app:eq:analytical_zmag_final}
    m^{(1)}_\parallel(t) = -\Z + \frac{2}{h_\parallel} \int_0^\pi \dpi{k} \vert K(k,t) \vert^2 \omega(k) + \Z \left(\int_0^\pi \dpi{q} K(q,t) \int_0^\pi \dpi{k} K(k,t) F(\vert q,-q,k,-k) +\text{h.c.} \right)\,.
\end{equation}
For small $h_\perp$ and $h_\parallel$, the double integral is negligible and \eqref{app:eq:analytical_zmag_final} reduces further to
\begin{equation} \label{app:eq:analytical_zmag_simple}
    m^{(1)}_\parallel(t) \approx -\Z + \frac{2}{h_\parallel} \int_0^\pi \dpi{k} \vert K(k,t) \vert^2 \omega(k)\,. 
\end{equation} 
A comparison of $m^{(1)}_\parallel(t)$ from \eqref{app:eq:analytical_zmag_final} with MPS data is shown in figure \ref{app:fig:zmag} for $h_\perp=0.2$ and $h_\parallel=0.02$, and for various other parameters in \ref{app:fig:many_zmag}. Note that \eqref{app:eq:analytical_zmag_final}/\eqref{app:eq:analytical_zmag_simple} agrees much better with the numerical results than the first order expansion of \eqref{app:eq:Z_1order}, which already fails quantitatively for small $h_\perp$ at later times.

\end{appendix}
\end{document}